\documentclass[notoc,paper,letterpaper]{JHEP3}
\usepackage{amssymb} 
\usepackage{graphicx}     			
\newcommand{\be}{\begin{equation}} 
\newcommand{\ee}{\end{equation}} 
\newcommand{\bea}{\begin{eqnarray}} 
\newcommand{\eea}{\end{eqnarray}} 
\def\ie{{\it i.e.}} 
\def\eg{{\it e.g.}} 

\def\tr{{\rm tr}}
\def\del{{\partial}}
\def\til{\widetilde}
\def\hat{\widehat}
\def\tf#1#2{{\textstyle{\frac{#1}{#2}}}}
\def\vev#1{{\langle{#1}\rangle}} 
\def\Z{\mathbb{Z}} 
\def\C{\mathbb{C}} 
\def\U{{\rm u}} 
\def\SU{{\rm su}} 
\def\SO{{\rm so}} 
\def\SL{{\rm sl}} 
\def\Sp{{\rm sp}} 
 
\def\Im{\mathop{\rm Im}}
\def\Re{\mathop{\rm Re}}
\def\hra{\hookrightarrow}
\def\with{\mathop{\,\textstyle{\rm w/}\,}}
\def\SCFT{{\rm SCFT}}
\def\br{{\bf r}}
\def\adj{{\bf ad}}
\def\a{{\alpha}}
\def\b{{\beta}}
\def\g{{\gamma}}
\def\d{{\delta}}

\def\th{{\theta}}
\def\l{{\lambda}}
\def\m{{\mu}}
\def\n{{\nu}}

\def\t{{\tau}}
\def\o{{\omega}}
\def\tT{{\til T_2}}
\def\tj{{\tilde\jmath}}

\title{S-duality in N=2 supersymmetric gauge theories}
\author{Philip C. Argyres$^1$ and Nathan Seiberg$^2$\\ 
$^1$\,Physics Department, University of Cincinnati, 
Cincinnati OH 45221-0011\\
\email{argyres@physics.uc.edu}
\vskip5pt
$^2$\,School of Natural Sciences, Institute for Advanced Study, 
Princeton NJ 08540\\
\email{seiberg@ias.edu}}

\abstract{A solution to the infinite coupling problem for  
$N=2$ conformal supersymmetric gauge theories in four dimensions 
is presented.
The infinitely-coupled theories are argued to be interacting
superconformal field theories (SCFTs) with weakly gauged flavor groups.  
Consistency checks of this proposal are found by examining some 
low-rank examples.  As part of these checks, we show how to compute
new exact quantities in these SCFTs: the central charges of their flavor 
current algebras.  Also, the isolated rank 1 $E_6$ and $E_7$ SCFTs 
are found as limits of Lagrangian field theories.}
 
\begin{document}

\section{Infinite coupling and S-duality}

In many $N=2$ supersymmetric gauge theories in four dimensions,
an exactly marginal gauge coupling, $g$, can be taken infinite. 
In this paper we propose a new kind of quantum equivalence of 
gauge theories which relates such infinitely-strongly coupled
theories to ones with both weakly-coupled ($g\ll1$) and 
strongly-coupled ($g\sim 1$) sectors, but no infinitely-strongly
coupled sectors.  This proposal thus allows one to eliminate
infinitely-coupled gauge theories in favor of merely strongly-coupled 
ones.  In particular, it suggests that even as $g \to \infty$ all 
the correlation functions of the theory remain finite!
Our proposal generalizes the well-known S-duality of $N=4$ supersymmetric
gauge theories to the larger class of $N=2$ supersymmetric ones.

\FIGURE{
\includegraphics[width=16em]{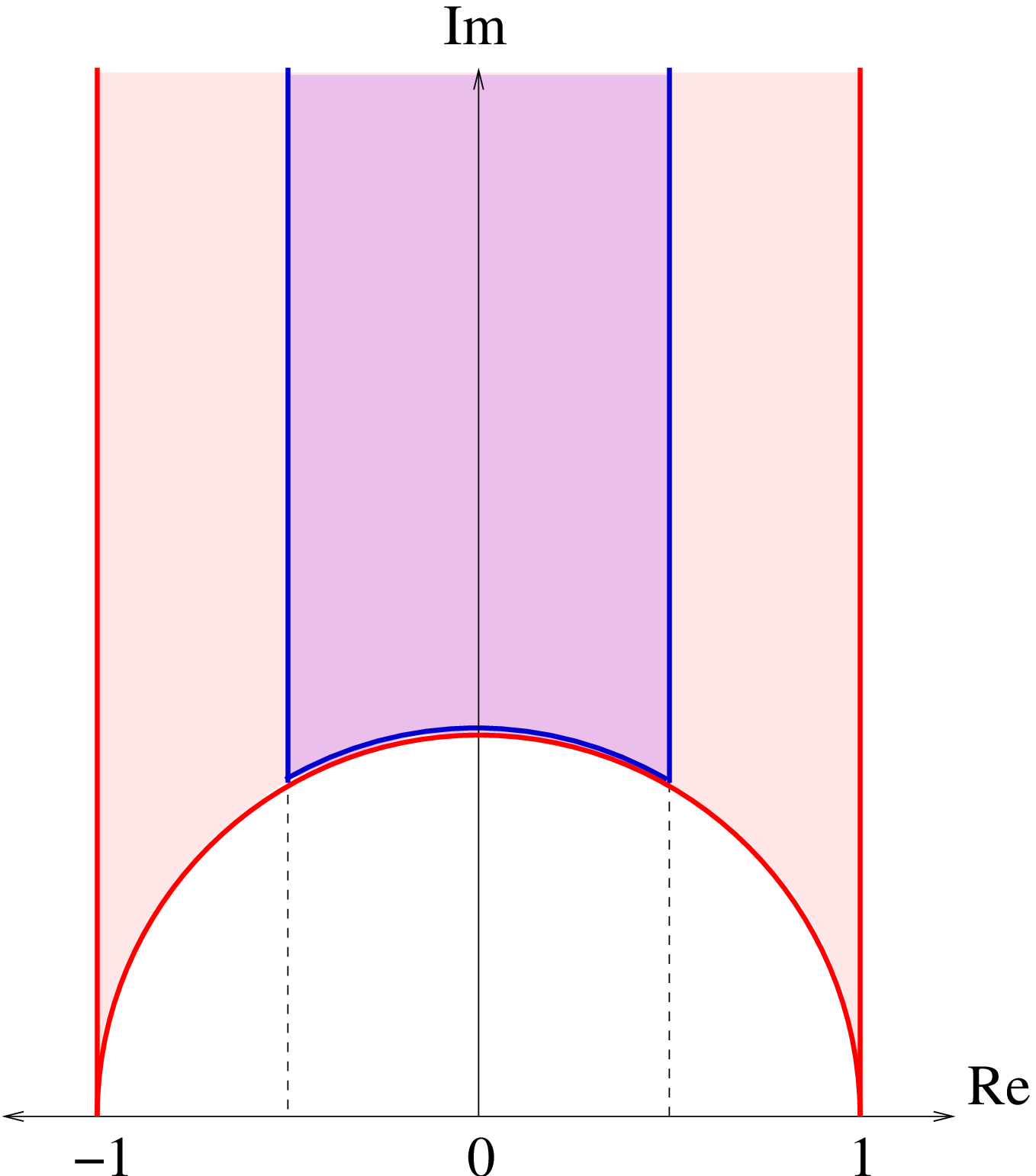}    
\caption{Fundamental domain in $\t$ for $\SL(2,\Z)$ (blue), and in 
$\til\t$ for $\Gamma^0(2)$ (red).  Edges of the domains are identified 
under reflection through $\Re\t=0$.}
}

S-duality, or Olive-Montonen duality \cite{OMduality}, in $N=4$ 
supersymmetric gauge theories in four dimensions answers the question 
of what happens as the gauge coupling constant becomes infinite:  the 
theory actually becomes a weakly coupled gauge theory again, though not 
necessarily with the same gauge group.  In theories with simply-laced
gauge group, where the theory is self-dual, this is expressed as the 
equivalence between the theory at different couplings, $\t\simeq-1/\t$,
where $\t=\th/2\pi + 4\pi i/g^2$ is the complex coupling.  Combined with 
the angularity of the theta angle, $\t\simeq\t+1$, this generates an 
$\SL(2,\Z)$ group of identifications whose fundamental domain in the space 
of couplings is bounded away from infinite coupling ($\Im\t=0$); see 
figure 1.

But this is not always the answer to the infinite coupling problem in 
scale-invariant gauge theories with less supersymmetry.  Though in the 
case of $N=2$ $\SU(2)$ superQCD with four massless fundamental 
hypermultiplets there is an $\SL(2,\Z)$ S-duality \cite{sw9408}, there 
are higher-rank gauge theories which are not self-dual.  For example, 
for $N=2$ $\SU(3)$ with six massless fundamentals, the S-duality group
is $\Gamma^0(2)\subset\SL(2,\Z)$ generated by \cite{aps9505} $\til\t
\simeq\til\t+2$ and $\til\t\simeq-1/\til\t$ where $\til\t:=2\t$.  The
fundamental domain of this group in the coupling space is not bounded 
away from infinite coupling, but instead contains the point $\til\t=1$, 
as shown in figure 1.

This raises the question of how to characterize the physics at the
infinite coupling point.  The simplest possibility is that this limit 
is actually a different weakly-coupled $N=2$ gauge theory, but the exact
low energy effective action shows that this cannot be the case.  One
way to see this is to compare the behavior of the curve 
\cite{sw9408} encoding the Coulomb branch effective action in the weak 
and infinite coupling limits.  As $\Im \til\t\to\infty$ in the $\SU(3)$ 
theory, three non-intersecting cycles pinch in the genus 2 curve at any 
point on the moduli space; see figure 2a. (More precisely, we are picking 
out the particular vanishing cycles for which there exist BPS states in 
the spectrum.)  Each pinching cycle is the signature of a pair of charged 
$W^\pm$ gauge bosons becoming massless, corresponding to the expected 6 
gauge bosons given a mass by the $\SU(3)\to\U(1)\times\U(1)$ Higgsing.  
In contrast, as $\til\t\to1$, the infinite coupling point, only one 
cycle vanishes at generic points on the moduli space.  This does not 
give enough $W$ bosons to account for a weakly-coupled Higgs mechanism.  
So, some new phase is indicated for this $N=2$ $\SU(3)$ gauge theory at 
$\til\t=1$.  This type of behavior of the Coulomb branch effective action 
is typical of the infinite-coupling points of many infinite series of 
such theories \cite{ab98}.  
\FIGURE{
\includegraphics[width=18em]{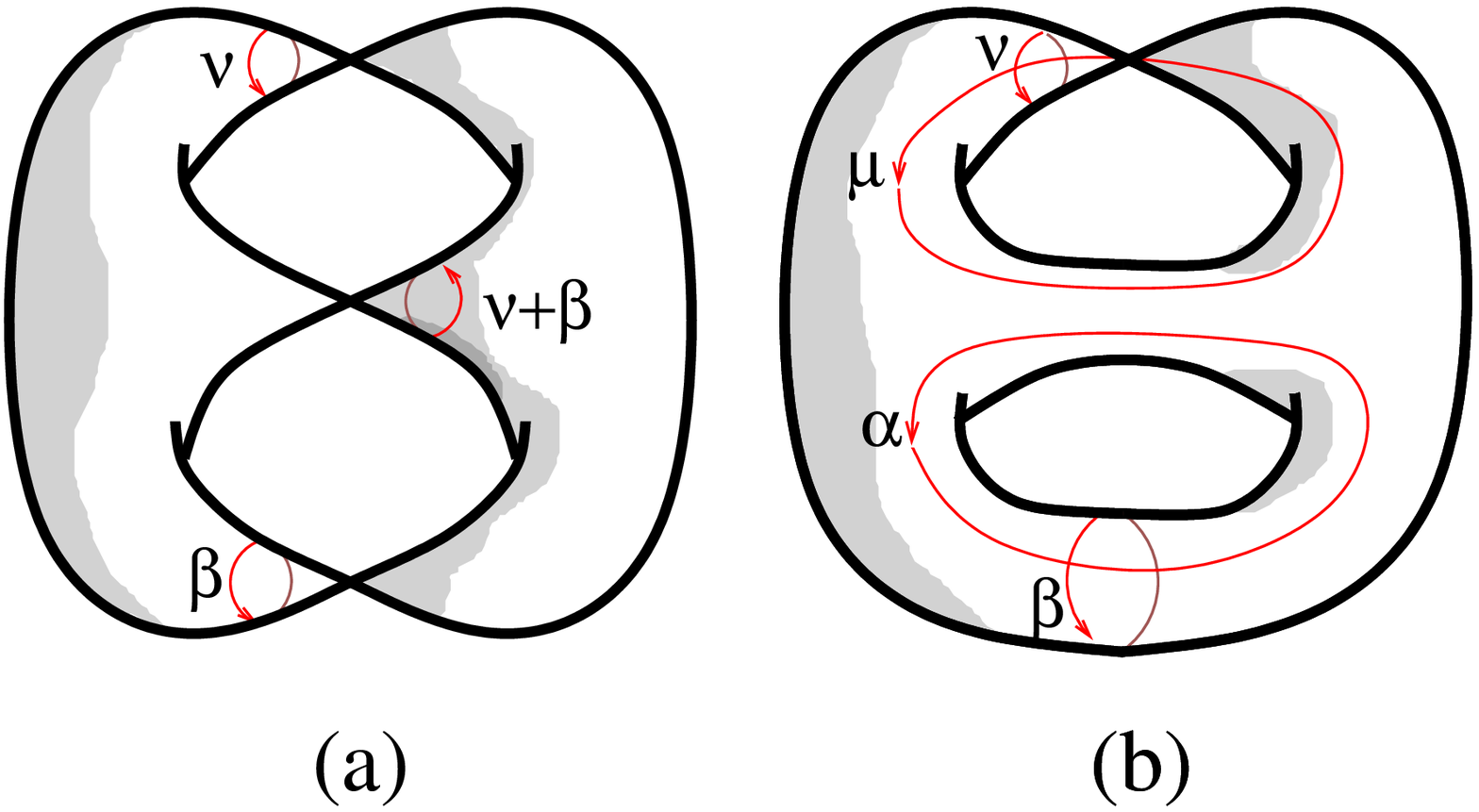}    
\caption{Degenerations of the scale invariant $SU(3)$ curve 
at a generic point on the Coulomb branch at (a) weak coupling, 
and (b) infinite coupling.}}

To avoid confusion, we should note that
the numerical value of $\Im\til\t$ is not really indicative of whether a
point in coupling space is such an infinite-coupling point or
not.  For, lacking any other non-perturbative definition of 
the coupling, one could always make a holomorphic, non-perturbative
redefinition of the coupling to change its value at the
putative infinite-coupling point to any desired value.
Indeed, in the rest of the paper
we will find it convenient to describe couplings in terms
of a function $f(\til\t)$ which approaches $f\sim e^{i\pi\til\t}
\to 0$ at weak coupling, but $f\to1$ at ``infinite coupling".

We nevertheless use ``infinite coupling" as a convenient phrase 
to describe those points in coupling
space where the effective action has the singular (but not weak-coupling)
behavior at generic points on the Coulomb branch described above.  
These infinite coupling points can be invariantly characterized as
follows.  Abstractly, the space of couplings is a complex manifold 
with singularities, and the S-duality group is the fundamental group 
in the orbifold sense \cite{w9703} of this space with the singular 
points removed.  The infinite coupling points we are interested in 
are the cusps, \ie\ the points where the S-duality identification is 
of infinite order (like the $\t\to\t+1$ theta angle identification 
at weak coupling).  The coupling is just a complex coordinate on this 
space.  If the coupling $\t$ transforms by fractional linear 
transformations under the S-duality group, then $g^2 \sim (\Im\t)^{-1}$ 
is either 0 or infinite at cusps. 

In this paper we argue that the physics at the infinite coupling limit 
of a scale-invariant $N=2$ gauge theory with gauge group $H$, rank$(H)=r$, 
is a {\em weakly coupled} scale-invariant gauge theory with gauge group 
$G$ with smaller rank, rank$(G)=s<r$, which is coupled to an isolated 
rank $(r-s)$ $N=2$ superconformal field theory.\footnote{The rank of an 
isolated $N=2$ SCFT, where an explicit gauge group and Lagrangian 
description are lacking, is defined as the complex dimension of its 
Coulomb branch, which is the number of $\U(1)$ gauge factors generically 
unbroken in the IR.}  Here ``isolated" means that this SCFT has no exactly 
marginal coupling of its own.  Thus, the SCFT can be thought of as 
providing ``matter fields" charged under $G$.  More precisely, in the 
infinite coupling limit, $G$ weakly gauges a subgroup of the flavor 
symmetry of the SCFT.  (We will always use ``flavor symmetry" to refer 
to the global symmetry which commutes with the $N=2$ superconformal 
symmetry.)

In other words, we are proposing a generalization of S-duality from 
$N=4$ to $N=2$ scale-invariant field theories.  Our proposal 
of including strongly-coupled $N=2$ SCFTs as factors in the duals of 
non-Abelian gauge theories is a natural generalization of the $N=4$ case, 
given the existence of isolated $N=2$ conformal gauge theories.

Let us describe the infinite-coupling duality more precisely.
Denote a gauge theory with gauge group $G$ and matter 
half-hypermultiplets in the $\bigoplus_i\br_i$ representation by
\be
G \with \bigoplus_i \br_i ,
\ee
where $\with$ is read as ``with".
Now consider a theory described by a SCFT with flavor 
symmetry group $S$, a subgroup $G\subset S$ of which is
gauged.  We will denote such a theory
in a similar manner as
\be
G \with \SCFT_S,
\ee
emphasizing that the SCFT acts as ``matter" for the gauge
theory (even though $N=2$ SCFTs also have gauge degrees of 
freedom).  Note that $S$ is {\em not} the flavor symmetry 
of this theory, except in the limit of zero coupling; at 
finite coupling only the subgroup of $S$ commuting with $G$ 
is a global symmetry.  In general there may be more than one 
way of embedding $G$ in $S$.  A given embedding can often be 
specified by the maximal subgroup $G\times F\subset S$.  
Gauging $G$ leaves $F$ as the flavor group.

In the rest of this paper we will give evidence for specific examples 
of the duality between infinitely-coupled Lagrangian $N=2$ SCFTs and 
gaugings of isolated SCFTs, as described above.  For example, in the
next section we argue that 
\be\label{e6duality}
\SU(3) \with 6\cdot({\bf3} \oplus \bar{\bf3})
=
\SU(2) \with \left( 2\cdot{\bf 2} \oplus \SCFT_{E_6} \right).
\ee
In words and more detail: the scale-invariant theory with gauge 
group $G=\SU(3)$ coupled to 6 massless fundamental hypermultiplets 
with coupling $f$ (reviewed in appendix A) is equivalent to an 
$\SU(2)$ gauge theory with one massless fundamental hypermultiplet 
and coupled to the isolated rank 1 SCFT with flavor symmetry $E_6$ 
(reviewed in appendix B) by gauging the $\SU(2)$ in 
the maximal subgroup $\SU(2)\times\SU(6)\subset E_6$ with coupling 
$\til f$.  The map between the direct and dual couplings, $\til f(f)$, 
is given in (\ref{cplgmap}), and maps infinite coupling in $f$ to 
zero coupling in $\til f$.

Two immediate checks of this proposal are that the ranks and flavor
groups of the two sides of (\ref{e6duality}) match.
The rank (or dimension of the Coulomb branch) of the $\SU(3)$ 
theory is 2, while the $\SU(2)$ and $E_6$ SCFT factors on the right
are each rank 1.  The flavor groups match since that on the left is 
$\U(6)$, while the $2\cdot{\bf2}$ factor on the right contributes a
$\U(1)$ and the $E_6$ SCFT contributes a $\SU(6)$ because of the 
way the gauged $\SU(2)$ factor is embedded in $E_6$.  This embedding
must also be consistent with the low energy effective action, giving
independent evidence for this proposal, described in section 2.

It is less trivial to see that the number of marginal couplings is the 
same on each side of (\ref{e6duality}).  Clearly there is only one
marginal coupling on the left side, but the $\SU(2)$ coupling on the
right will be marginal only if the contribution of the rank 1 SCFT 
``matter" to the $\SU(2)$ gauge coupling beta function has the
correct value.  This contribution is governed by the central charge 
of the flavor current algebra of the rank 1 SCFT.  By
weakly gauging the global flavor symmetries on both
sides of the duality and comparing its gauge coupling beta functions
we can independently compute this central charge, and verify
that the $\SU(2)$ gauge coupling is indeed marginal.
This is described in detail in section 3, where the
same argument is used to give evidence that 
the $\SU(2)$ $W^\pm$-bosons are magnetically charged under the $\SU(3)$ 
gauge group, indicating that the $\SU(2)$ gauge group on the 
right side of (\ref{e6duality}) is {\em not} a subgroup
of the $\SU(3)$ gauge group on the left side.

An important result of this paper is the above computation
of the flavor current algebra central charge of the SCFT.  
This is a new exactly computed observable of isolated, 
strongly-coupled $N=2$ SCFTs.

A somewhat simpler example is outlined in section 4, where it is
shown that
\be\label{e7duality}
\Sp(2)\with 12\cdot{\bf 4} =  \SU(2) \with \SCFT_{E_7},
\ee
with the $\SU(2)$ gauging the $\SU(2)$ factor in the
maximal embedding $E_7 \supset \SU(2)\times\SO(12)$, 
to realize the $\SO(12)$ flavor symmetry.

It is worth noting that the conjectures (\ref{e6duality}) and
(\ref{e7duality}) identify the $E_6$ and $E_7$ rank 1 SCFTs as 
subsectors of Lagrangian field theories which decouple from the 
rest of the theory in the infinite coupling limit.  The rank 1 
$N=2$ SCFTs with exceptional global symmetry groups \cite{mn9608} 
have not been previously constructed in a purely four-dimensional 
field theory framework; instead they have been shown to exist only 
by dimensional reduction from a 5 or 6 dimensional SCFT, which in 
turn were constructed as low energy limits of certain string 
configurations \cite{excft}.

Other scale invariant rank 2 theories (listed in appendix A.1) 
have infinite coupling limits whose dual descriptions can be analyzed 
similarly.  We leave this for later, though, since these examples require 
substantially more work because less complete information is available 
either about the Coulomb branch effective actions of these theories or 
about their proposed dual SCFTs.  In particular, verification of the
same consistency checks as performed for the two examples (\ref{e6duality})
and (\ref{e7duality}) in this paper requires the 
computation of non-maximal mass deformations of the $E_n$ rank 1 SCFTs 
and of the mass-deformed curve of the $G_2 \with 8\cdot\bf7$ rank 2 
Lagrangian SCFT.
The non-maximal mass deformations are mass deformations of SCFTs which
have the same conformal curves as the $E_n$ SCFTs, but have smaller
global symmetries than the maximal deformations 
with $E_n$ symmetry.  Some other examples of identical conformal
curves which have inequivalent mass deformations are pointed out
in appendix A.

More generally, infinite coupling points are ubiquitous in higher-rank 
$N=2$ Lagrangian theories.  For example, all the $n>2$ $\SU(n)$ 
theories with $2n$ fundamental hypermultiplets have 
infinite coupling limits, and similarly for many other series of 
scale-invariant theories with rank greater than 2.  In fact, the only 
series of higher-rank theories which are known to have Lagrangian weak 
coupling descriptions at all the singularities in the space of marginal 
couplings, besides the $N=4$ theories, are the $\Sp(n)$ theories with 4 
fundamentals and 1 antisymmetric \cite{asty9611,dls9612}, of which 
$\SU(2)$ with four fundamentals \cite{sw9408} is a special case, and 
the $\SU(2)\times\SU(2)$ theory with 4 fundamentals and 1 bifundamental
\cite{a9706,ab9910}.  However, it is probable that all the $\SU(2)^n$ 
cylindrical and elliptic models \cite{w9703} are also examples of this type.

Finally, we have included several appendices in an attempt to make
the paper more self-contained.  They mostly either collect 
scattered results from the $N=2$ field theory literature, or
review results which are probably known to experts but have 
not been published.

\section{Infinite coupling in su(3) w/ 
$6$ $\cdot$ ($\bf{3\oplus\bar3}$)}

Our aim is to check the proposed equivalence 
\be\label{su3=e6}
\SU(3) \with 6\cdot({\bf3}\oplus{\bf\bar3}) 
= 
\SU(2) \with (2\cdot{\bf 2} \oplus \SCFT_{E_6}).
\ee
In this section we extract evidence supporting it from
the known low energy effective action on the Coulomb branch
for the $\SU(3)$ theory.

\subsection{Limits in the effective theory}

As reviewed in appendix A.2, the curve encoding the low 
energy effective action on the Coulomb branch of the
$\SU(3)$ theory with six massless fundamental hypermultiplets 
is (\ref{A2+6.3}).  $u$ and $v$ are the Coulomb  branch vevs of 
dimension 2 and 3, respectively, and their associated basis of 
holomorphic one-forms are given in (\ref{basis}).  These one-forms
determine the central charges and thus the 
masses of BPS states.  The infinite coupling 
point is at coupling $f=1$.  The curve can be conveniently factorized as
\be\label{su3fac}
y^2 = \left[ (1-\sqrt f)x^3 -ux-v\right]
\left[ (1+\sqrt f) x^3 -ux -v\right] .
\ee
As $f\to1$, this factorization makes it clear that the curve degenerates 
to a genus one curve, and it is easy to check that the $\o_u = xdx/y$ 
one-form develops a pair of poles at $x=\infty$, whereas $\o_v = dx/y$ 
remains holomorphic.  Figure 2b is a representation of this degenerate 
curve.

\FIGURE{
\includegraphics[width=12em]{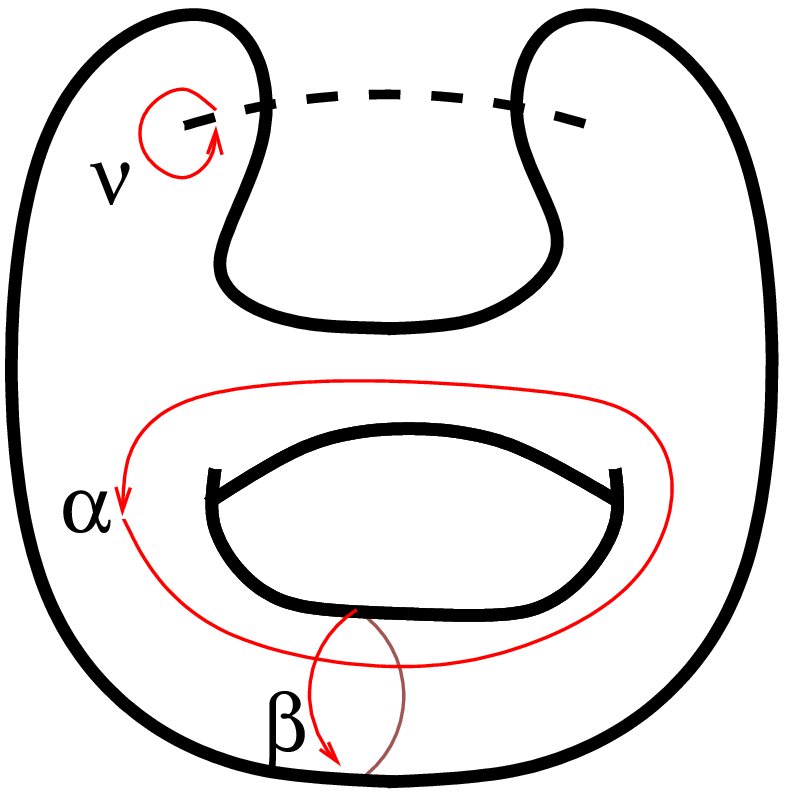}    
\caption{Curve of a rank one theory with massive
quark.  The dotted line marks the pinched handle 
that unpinches upon weakly gauging the quark number 
symmetry.}}

This degeneration corresponds to what one expects the curve to
be in the weak coupling limit of the right side of (\ref{su3=e6}),
as we now explain.  Recall that mass parameters in $N=2$ theories 
transform in the adjoint representation of the associated global 
flavor symmetry which they break.  A rank one theory with a Coulomb 
branch vev $u$ and a single mass parameter $m$ has an associated 
rank 1 global flavor symmetry (``quark number").  Its central charge 
has the form \cite{sw9408} $Z= e a(u) + g a_D(u) + n m$, where $e$ 
and $g$ are the electric and magnetic charges of the low energy 
$\U(1)$ gauge group, and $n$ is the quark number.  Now imagine 
weakly gauging the quark number symmetry so that $n$ becomes the 
new electric charge of this gauge group and the mass $m$ becomes 
the vev of the associated vector multiplet.  Thus the central charge 
obeys $\del Z/\del u  = \oint_\g \o_u$ and $\del Z/\del m = \oint_\g 
\o_m$ for holomorphic one forms $\o_{u,m}$ and cycles $\g$ on the 
associated genus two curve.  The weaker this gauging, the heavier the 
magnetic monopoles of the quark number $\U(1)$ become,
corresponding to a cycle intersecting a vanishing 
cycle of the degenerating genus two curve, as in the $\mu$ cycle of 
figure 2b.  When the quark number gauging is turned off, the curve 
becomes a nondegenerate genus one curve with holomorphic one-form 
$\o_u$, as shown in figure 3.  From the form of the central charge, 
it follows in this limit that $\del Z/\del m = n = \oint_\g \o_m$ 
for $[\g] = e[\a] + g[\b] + n[\n]$, and therefore that the 
second one-form, $\o_m$, develops a pair of poles.

Thus, the degeneration of the 
$\SU(3)$ curve when $f=1$ may be interpreted as follows: $v$ 
becomes the Coulomb branch vev 
of a rank 1 SCFT in which $u$ appears as a mass 
deformation.  For $f$ close to but not equal to 1, $u$ is the
vev of a vector multiplet weakly gauging some global symmetry of
this rank 1 SCFT.

Furthermore, the curve of the rank 1 SCFT can be explicitly identified
by setting $u=0$ and $f=1$.  Then the curve (\ref{su3fac}) and
holomorphic one-form become
\be
y^2=-v(2x^3-v), \qquad \o_v = dx/y .
\ee
Making the change of variables $x = -i\til x/\til v$, $y=2\til y/\til v$,
$v=2i\til v$, this becomes
\be\label{su3e6}
\til y^2 = \til x^3 - \til v^4, \qquad
\til\o_{\til v} = d\til x/\til y ,
\ee
which is the curve of the rank 1 $E_6$ SCFT, reviewed in appendix B.

Also, the curve of the $\SU(2)$ factor which weakly gauges part of 
the flavor symmetry of the $E_6$ SCFT can be extracted by setting
$v=0$ (the conformal point of the $E_6$ theory) in (\ref{su3fac}),
\be
y^2 = x^2 \left[ (x^2-u)^2 - f x^4 \right],  \qquad
\o_u = xdx/y .
\ee
The factor of $x^2$ is due to a pinched cycle at $x=0$.
Scaling to the remaining genus one curve by defining $\hat y = y/x$
gives the curve
\be\label{su2curve}
\hat y^2 = (x^2-u)^2 - f x^4, \qquad \o_u = dx/\hat y ,
\ee
which is precisely the curve \cite{aps9505} of the scale
invariant $\SU(2)$ $N=2$ superQCD.  It is weakly coupled
at $f=0$ and $f=1$, these limits being related by the $\SL(2,\Z)$
S-duality of this theory \cite{sw9408}.  In particular,
this S duality maps the the curve (\ref{su2curve}) to itself
with $f\to\til f(f)$ where $\til f(1)=0$.
Indeed, this strong to weak coupling map can be
extracted from \cite{aps9505}: 
\be\label{cplgmap}
\til f 
= 8{(1-f)(4-9f)+(4-3f)\sqrt{1-f} \over (8-9f)^2} .
\ee
This is an exact relation for the couplings implicitly
defined by how they appear in (\ref{su2curve}) and
its dual.  In the weak coupling limit, this can be
related to the traditional gauge coupling defined by
$q = e^{2\pi i\tau} = e^{i\th} e^{-8\pi^2/g^2}$.
For the $\SU(2)$ and $\SU(3)$ Lagrangian SCFTs
with curves (\ref{su3fac}) and (\ref{su2curve}),
$f\approx -64 q$ when $f$ is near 0 \cite{aps9505}.  
Thus (\ref{cplgmap}) implies that near the infinite
coupling point, the dual coupling of the $\SU(2)$ 
factor goes as $\til q \approx -(1/2)\sqrt{1-f}$.

Note that the elements of this duality---that the infinite 
coupling dual involves the $\SU(2)$ and $E_6$ SCFTs---could 
have been guessed just from the anomalous dimensions of the 
chiral operators on the Coulomb branch: $u$ and $v$ have
dimensions 2 and 3, while the $E_6$ SCFT is the only rank 1 SCFT 
with vev of dimension $3$, and the $\SU(2)$ SCFT is likewise
the unique rank 1 SCFT with vev of dimension 2 \cite{mn9608}.  

\subsection{Global symmetries}

The $\SU(3)$ theory with six massless fundamental hypermultiplets
has global symmetry group $\SU(6)\times\U(1)\times\U(2)_R$.
The $\U(2)_R$ R-symmetry is part of the $N=2$ superconformal
algebra, so is also automatically present in the proposed
$\SU(2)$ product $E_6$ SCFT at the infinite coupling point.
We now examine how the $\SU(6)\times\U(1)$ flavor
symmetries are realized in the dual theory.

\paragraph{The su(6) symmetry.}  

To be consistent with our proposal that $\SU(3)$ with 6 fundamentals 
near infinite coupling is the $E_6$ 
rank 1 SCFT with the $\SU(2)$ factor of the maximal $\SU(2) \times
\SU(6)\subset E_6$ subgroup weakly gauged, it follows that as 
the $u$ vev of the $\SU(2)$ factor is turned on, this maximal 
subgroup of $E_6$ must break as
\be\label{su3-breaking2}
\SU(2)\times\SU(6) \to 
\U(1)\times\SU(6) .
\ee
In terms of the
eigenvalues $m_i$ ($i=1,\ldots,6$) and $m$ of the Cartan subalgebra 
of the $\SU(6)\times\SU(2)$ maximal subgroup of $E_6$ introduced
in appendix B.1, the breaking (\ref{su3-breaking2}) corresponds to
\be\label{su3-breaking3}
m \propto \sqrt{u},\ \ m_{1,\ldots,6}=0.
\ee

This can be checked from the low energy effective action as follows.
At $f=1$, turn on $u$ to get the curve
\be
y^2 = -(ux+v)(2x^3-ux-v), \qquad \o_v = dx/y .
\ee
Change variables according to the combined $\SL(2,\C)$ transformation
and rescalings
\be\label{su3-sl2c}
x = {-2i\til v(u^2 + 12\til x)\over u^3+24\til v^2 + 12u\til x},\qquad
y = {-1152\til v^3\til y\over(u^3+24\til v^2 + 12u\til x)^2},\qquad
v = 2i\til v,
\ee
to bring this into the form 
\be\label{su2-e6def}
\til y^2 = \til x^3 - \left(\til v^2 u + {1\over48} u^4 \right) \til x
- \left( \til v^4 + {1\over12} \til v^2 u^3 + {1\over 864} u^6 \right),
\qquad \til\o_{\til v} = {d\til x\over\til y} .
\ee
This is a deformed version of the $E_6$ SCFT curve (\ref{su3e6}).
The general mass deformation of
the $E_6$ SCFT is given by (\ref{e6-e6def}).  Comparing to 
(\ref{su2-e6def}) gives the $E_6$ adjoint casimirs 
\be\label{su3-breaking}
M_2 = u , \qquad
M_5 = 0 , \qquad
M_6 = {u^3\over12} , \qquad
M_8 = {u^4\over48} , \qquad
M_9 = 0 , \qquad
M_{12} = {u^6\over864} .
\ee
Note that these coefficients
are ambiguous only up to the choice of normalization
of $u$.  Minahan and Nemeschansky \cite{mn9608}
determined the $M_n$ in terms of an explicit basis of $E_6$ 
Casimirs.  In appendix B.1 we have rewritten these $E_6$ Casimirs 
in terms of the Casimirs of its $\SU(2)\times\SU(6)$ 
maximal subgroup.  Compare (\ref{su3-breaking}) to 
(\ref{e6crv-e6}), (\ref{e6-casimirs}), and (\ref{su6su2-casimirs})
to identify how turning on $u$ breaks the $E_6$ global
symmetry.  One easily checks that the assignments $m=\sqrt{u/2}$ 
and $m_i=0$, consistent with (\ref{su3-breaking3}), reproduce
(\ref{su3-breaking}).\footnote{There are a few subtleties in
performing this matching.  Some algebra shows that (\ref{su3-breaking}) 
is actually consistent with three different mass assignments (up to permutations),
$$
\begin{array}{rcrcl}
\mbox{(a):} && m=\sqrt{u/2},\ \ m_{1,\ldots,6}=0,
&\ \Longleftrightarrow\ &
{\bf su(2)}\times\SU(6) \to 
\ \, {\bf u(1)}\times\SU(6) ,\\
\mbox{(b):} && m=m_{1,\ldots,4}=0,\ \ m_5 = -m_6 = \sqrt{u/2},
&\Longleftrightarrow&
{\bf su(2)}\times\SU(6) \to 
{\bf su(2)}\times\SU(4)\times\U(1)^2 ,\\
\mbox{(c):} && m=m_{1,2,3}=\sqrt{u/8},\ \ m_{4,5,6}=-\sqrt{u/8},
&\Longleftrightarrow&
{\bf su(2)}\times\SU(6) \to 
\ \, {\bf u(1)}\times\SU(3)^2\times\U(1) ,
\end{array}
$$
where the corresponding adjoint breaking patterns
of the $\SU(2)\times\SU(6)$ maximal subgroup of $E_6$ are shown
on the right.
The (a) breaking, which manifestly leaves an unbroken $\SU(6)$ factor,
is the one described above.
The (b) breaking pattern actually gives the same picture, since it
also leaves an $\U(1)\times\SU(6)\subset E_6$ unbroken.  This is not
manifest because the unbroken $\SU(6)$
does not coincide with the $\SU(6)$ used for the basis of Casimirs.
(There are three inequivalent ways of embedding $\SU(2)\times\SU(6)$
in $E_6$, related by triality of the affine $\widehat E_6$ root system.
Two of these are related by complex conjugation in $E_6$, and so 
give the same adjoint breaking, the (b) pattern; the third is the
(a) pattern.)  In particular, in the (b) breaking, the unbroken
$\SU(6)$ factor is realized as $\SU(2)\times\SU(4)\times\U(1)$.
The (c) breaking pattern is different, and corresponds to adjoint
breaking of one $\SU(3)$ factor in the $\SU(3)^3\subset E_6$
maximal subgroup.  This does not give the expected global
symmetry group of the original $\SU(3)$ SCFT, but is ruled out
by the next check.}

A further test of this embedding of the $\SU(6)$ flavor symmetry in
$E_6$ can be realized by not only turning on the
$u$ vev at $f=1$, but also the fundamental masses
$m_i$ in the $\SU(3)$ theory.  From (\ref{A2+6.3massive})
the curve at $f=1$ is
\be\label{su3-massive}
y^2 =  - (2u + S_2) x^4 - (2v - S_3) x^3 
+ (u^2 - S_4) x^2 + (2uv + S_5) x 
+ (v^2 - S_6)
\ee
where the $S_n$ are the $\SU(6)$ Casimirs introduced in
the appendix.
The infinite coupling equivalence implies there should exist
a change of variables as in (\ref{su3-sl2c}) to bring this
to the form of the mass deformed $E_6$ SCFT curve (\ref{e6-e6def}).
It is too difficult to find this change of variables explicitly,
but we can find evidence that it exists by taking the discriminants of
the right sides of (\ref{su3-massive}) and (\ref{e6-e6def})
with respect to $x$, and comparing.  One finds that the two 
discriminants indeed agree with the identifications
\be
v = 2i\til v + 2i\til S_3 ,
\qquad
u = 2\til T + 4 \til S_2 ,
\qquad
S_n = (-2i)^n\til S_n ,
\ee
which fix, in addition to the necessary shifts and rescalings
of the $\SU(3)$ vevs $(u,v)$ relative to the masses and vev of
the $E_6$ curve, a rescaling of the $\SU(6)$ mass
eigenvalues between the two curves by a factor of $-2i$.
The $2i$ rescaling of the masses agrees with the one-form 
rescaling $\til\o_{\til v} = (\del v/\del\til v) \o_v
= 2i \o_v$, since the one-form normalization determines that
of the masses through (\ref{swform}), and since the $\SU(6)$ 
is an index one subgroup of $E_6$.

\paragraph{The u(1) symmetry.}

The $\U(1)$ factor of the flavor group of the $\SU(3)$ theory
on the left side of (\ref{su3=e6}) is realized on the right side 
as the $\SO(2)$ flavor symmetry rotating the two pseudoreal
half-hypermultiplets in the $\bf2$ of $\SU(2)$.  So this part of
flavor symmetry is realized in terms of weakly coupled degrees of
freedom at the infinite coupling point.  This is also indicated in
the effective action, since the curve (\ref{A2+6.3massive})
for the $\SU(3)$ theory with 6 fundamental hypermultiplets (with 
the $\SU(6)$ masses set to zero) is
\be
y^2 = \left[ (x+\sqrt{1-f}\,M)^3 - u (x+\sqrt{1-f}\,M) -v
\right]^2 -f x^6,
\ee
from which it is apparent that the $\U(1)$ mass deformation
vanishes at $f=1$.  This behavior is like that at weak coupling
where all the mass deformations vanish, and is in accord with
the $\U(1)$ flavor symmetry being associated to matter
charged only under the $\SU(2)$ gauge group which is weakly
coupled at $f=1$.

\section{Beta functions and central charges}

The proposed duality requires the $\SU(2)$
gauge factor on the right side of (\ref{su3=e6}) to be 
scale invariant.  This means that the $2\cdot{\bf2} \oplus 
\SCFT_{E_6}$ ``matter" must contribute just enough 
to the beta function for the $\SU(2)$ gauge coupling 
$\til f$ to cancel the contribution from the adjoint 
$\SU(2)$ vector multiplet.  The contribution from the $2\cdot
{\bf2}$ half-hypermultiplets follows from the standard perturbative 
computation (one-loop exact by a non-renormalization theorem), 
but the contribution from the rank 1 $E_6$ SCFT does not, since 
the $E_6$ theory is a strongly coupled theory.  Thus demanding
the vanishing of the beta function allows us to compute the
$E_6$ SCFT contribution.  

We do this in section 3.1, where we also relate this beta function
contribution to the central charge, $k$, of the $E_6$ current 
algebra.  They are related because the beta function is
proportional to the 2-point function of the gauge currents, 
and the gauge currents are linear combinations of the $E_6$ 
currents since the $\SU(2)$ gauge group is a subgroup of the
$E_6$ flavor group of the SCFT.
The central charge $k$ is analogous to the
current algebra central charge familiar from 2-d CFTs, and is a 
new exactly computable observable of these 4-d SCFTs.

In section 3.2 we give a different way of using the proposed
duality to compute $k$.  In a spirit similar to that of 't Hooft's
anomaly matching argument \cite{tHooft}, one can weakly gauge the 
flavor symmetry and compute the contribution to its beta function on
both sides of the duality.  Since at infinite coupling the $\SU(6)$
flavor symmetry is also a subgroup of the $E_6$ SCFT flavor group,
the contribution to its beta function also depends on $k$.
The agreement of this calculation with the previous one
is a non-trivial check of the duality.

We also apply this weak-gauging argument to the $\U(1)$ flavor
symmetry to find evidence that the gauge bosons of the $\SU(2)$
dual gauge group are magnetically charged with respect to
the original $\SU(3)$ gauge group.  Finally, we comment on the
application of this argument to the $R$-symmetries as well.

\subsection{Gauge coupling beta function at infinite coupling}

The effective gauge coupling is the coefficient of a
term quadratic in the gauge fields, so can be computed 
in a background field formalism by a two point function for the 
background gauge bosons.  The contribution to this correlator 
from the matter charged under the gauge group is then proportional 
to the two point function of the 
conserved current to which the gauge bosons couple.
Lorentz and scale invariance and current conservation imply
the OPE of currents $J^a_\mu$ for a (simple) symmetry group $G$ must
have the form
\be\label{flavorOPE}
J_\m^a (x) J_\n^b (0) =
{3k\over4\pi^4} \d^{ab} {x^2 g_{\m\n} - 2 x_\m x_\n \over x^{8
}} + {2\over\pi^2} f^{abc}
{ x_\m x_\n \, x\cdot J^c(0) \over x^{6
} } + \ldots
\ee
Here $k$, the central charge, is defined relative to the normalization 
of the structure constants $f^{abc}$ which are in turn fixed
in this paper by choosing the long roots of the Lie algebra to have
length $\sqrt2$.
The coefficient of the $f^{abc}$ term has been chosen so that 
$[Q^a,Q^b] = i f^{abc} Q^c$, where $Q^a:=\int d^3x J_0^a$ is the 
conserved flavor charge, as can be checked by appropriately integrating 
(\ref{flavorOPE}).  Also the factor of $3/4\pi^4$ in the central
charge term has been chosen to agree with the central charge 
normalization convention used in \cite{bgisw0507}.

As discussed in appendix D, the central charge for the $\U(n)$ flavor 
current algebra of $n$ free half-hypermultiplets is $k=1$, and the 
central charge for the half-hypermultiplet currents of a weakly gauged 
subgroup $G\subset\U(n)$ such that $\bf n = \oplus_i \br_i$ under $G$
is 
\be\label{tauhyper}
k_{G\rm -hypers} = \sum_i T(\br_i) ,
\ee
where $T(\br_i)$ is the quadratic index normalized as in appendix C.  
This is just the contribution to the beta function of the gauge coupling 
of the half-hypermultiplets.  Thus, the contribution to the central charge 
of the gauge current algebra by vector multiplets in a gauge group $G$ 
will be 
\be\label{tauvector}
k_{G\rm -vector} = -2T(\adj)
\ee 
in this normalization, in order to produce the known beta function, 
$-2T(\adj)+\sum_iT(\br_i)$, of the Lagrangian field theory $G\with
\oplus_i \br_i$.

Say an isolated SCFT has flavor symmetry $H$, and denote the central 
charge of its flavor current algebra by $k_H$.  Upon weakly gauging a 
subgroup of this flavor symmetry, $G\subset H$, the arguments of appendix 
D show that the contribution of the $H$ ``matter" to the gauge current 
central charge is
\be\label{tauSCFT}
k_{G\subset H} = I_{G\hra H}\, k_H,
\ee
where $I_{G\hra F}$ is the index of embedding of $G$ in $H$, 
defined in appendix C.

Putting (\ref{tauhyper}--\ref{tauSCFT}) together, the central charge for 
the $G$ gauge current algebra---proportional to the coefficient of 
the beta function for the gauge coupling---in the theory $G \with 
( \oplus_i\br_i ) \oplus \SCFT_H$ is $k_{G\rm -vector}+k_{G\rm -hypers}
+k_{G\subset H}=- 2 T(\adj) + \sum_i 
T(\br_i) + I_{G\hra H}\, k_H$.  For a scale invariant theory the beta 
function must vanish, giving the central charge of the isolated SCFT 
flavor algebra as
\be\label{CFTcentral}
k_H =  { 2 T(\adj) - \sum_i T(\br_i) \over I_{G\hra H} } .
\ee

Now apply this to our present example: $\SU(2)\with (2\cdot{\bf2}
\oplus \SCFT_{E_6})$.  For $\SU(2)$ in the normalization of 
appendix C, $T({\bf 3})=4$ and $T({\bf 2})=1$. 
Also, for $\SU(2)$ embedded in $E_6$ as the $\SU(2)$ factor of
the $\SU(2)\times\SU(6)$ maximal subalgebra, 
$I_{\SU(2)\hra E_6}=1$.  Thus we get
\be
k_{E_6} = {2 T({\bf 3}) - 2 T({\bf 2}) \over
I_{\SU(2)\hra E_6}} 
= 6.
\ee
This is a new, exactly computed observable in the strongly-coupled
rank 1 $E_6$ SCFT.

\subsection{Global symmetry central charges}

We can perform a check on this result by comparing the central charges
of the flavor current algebras at weak and infinite coupling.
They should be the same, since the value of the central charge cannot 
depend on the coupling.  One way to see that is as follows.  Above we 
viewed the mass parameters
as vevs of the scalar components of background vector multiplets which
gauge the flavor symmetry of the theory.  We can further explore these
gauge superfields and consider their one loop beta function.
The beta function for this flavor coupling
is, by a non-renormalization theorem, given exactly by its one-loop
contribution, which is independent of the value of the (original)
gauge coupling.  Thus the flavor beta function, and therefore the central
charge, should be the same at both weak and strong coupling.
This is similar in flavor to 't Hooft's anomaly matching argument
\cite{tHooft}, 
though the comparison here is being made between different values 
of a marginal coupling instead of between UV and IR scales.

The coupling indpendence of the central charge can also be seen
directly from the structure of representations of the $N=2$ 
superconformal algebra.
(Superconformal algebras and their unitary representations are
reviewed, for example, in \cite{minwalla}.)
Conserved flavor currents fall into 
$N=2$ superconformal multiplets whose primary has scaling 
dimension $D=2$, $\SU(2)_R$ spin $I=1$, $\SU(2)\times\til\SU(2)$ 
Lorentz spins $j=\tj=0$, and $\U(1)_R$ charge $R=0$.  
We denote this primary by $J^a_{ij}$ where $a$ labels the
generators of the flavor symmetry, while the symmetrized $ij$
are the $\SU(2)_R$ indices.  Thus the flavor current
central charge appears in the $\vev{J^a_{ij}(y) J^b_{k\ell}(z)}$
correlator.

On the other hand, the marginal gauge coupling $f$ multiplies
a chiral superfield term in the Lagrangian, $f \int d^4\th\,
\Phi$.  We see this at weak coupling, where $\Phi =\tr(W^2)$
and $W$ is the scalar chiral $N=2$ vector multiplet field strength
superfield.  Since terms in the Lagrangian are superconformally 
invariant, $\Phi$ must have $D=2$, $R=4$, and $I=j=\tj=0$.  Thus, 
the $f$-derivative of the flavor current algebra central charge
is measured by the three-point function:
$$
{\del k\over\del f} \sim \int d^4\th\, 
\vev{\Phi(x,\th) \, J^a_{ij}(y) \, J^b_{k\ell}(z)} .
$$
But in a superconformal theory, the vacuum expectation of a product
of superconformal primaries can only be non-zero if the total
$R$-charge of the primaries vanishes.  Since $R=4$ for $\Phi$,
but vanishes for $J^a_{ij}$, we conclude that $\del k/\del f=0$,
and therefore that the flavor current central charge is
independent of the marginal gauge coupling.

\paragraph{The su(6) symmetry.}

We start with the $\SU(6)$ flavor symmetry of the $\SU(3) \with 
6\cdot(\bf3\oplus\bar3)$ theory.  At weak coupling, the
vector multiplets are neutral under the flavor symmetry, and the
half-hypermultiplets transform as a $({\bf3},\bar{\bf6})\oplus
(\bar{\bf3},{\bf6})$ under the $\SU(3)\times\SU(6)$ combined
gauge and flavor symmetries.  Thus the central charge of the
flavor current is by (\ref{khypers})
\be\label{weaksu6}
k_{\SU(6){\rm -weak}} = 3\cdot T(\bar{\bf6})+3\cdot T({\bf6}) 
= 3\cdot 1+3\cdot1=6.
\ee
At infinite coupling the $\SU(6)$ flavor symmetry is realized as
the $\SU(6)$ factor of the $E_6\supset\SU(2)\times\SU(6)$
maximal subgroup for which $I_{\SU(6)\hra E_6}=1$,
from which it follows by (\ref{kGform2}) that
\be\label{strongsu6}
k_{\SU(6){\rm -strong}} = I_{\SU(6)\hra E_6} \, k_{E_6} = 6.
\ee
The agreement of (\ref{weaksu6}) and (\ref{strongsu6}) is
a non-trivial check on our proposal.

\paragraph{The u(1) symmetry.}  At weak coupling, the 
half-hypermultiplets in the $({\bf3},\bar{\bf6})$ of 
the $\SU(3)\times\SU(6)$ combined
gauge and flavor symmetries have charge $+1$ under the
$\U(1)$ flavor group, while those in the
$(\bar{\bf3},{\bf6})$ have charge $-1$.
(These charge assignments just amount to a choice of 
the normalization of the flavor $\U(1)$ generator.)
Thus, if this $\U(1)$ were weakly gauged, the matter 
multiplets would contribute to the coefficient of its 
beta function an amount proportional to the $\U(1)$
current algebra central charge
\be\label{weaku1}
k_{\U(1){\rm -weak}} = 3\cdot6\cdot(+1)^2 + 3\cdot6\cdot(-1)^2 = 36.
\ee
At infinite coupling the $\U(1)$ flavor symmetry is realized
as the $\SO(2)$ rotation symmetry of the two half-hypermultiplets
in the $\bf2$ of the $\SU(2)$ gauge group.  Then
one $\bf2$ has $\U(1)$ charge $+q$ and the other has
charge $-q$, but we can't determine $q$ {\it a priori} since
we don't have a direct way of comparing the normalization of 
the $\U(1)$ generator at infinite coupling and at weak coupling.
The $E_6$ SCFT ``matter" is not charged under the $\U(1)$.
Upon weakly gauging the $\U(1)$, the contribution to its
beta function from the matter multiplets will therefore be
\be\label{strongu1}
k_{\U(1){\rm -strong}} = 2\cdot(+q)^2 + 2\cdot(-q)^2 = 4q^2.
\ee
Equating (\ref{weaku1}) and (\ref{strongu1}) then implies $q=3$.

Although this value of the $\U(1)$
flavor charge at infinite coupling cannot be used as a 
consistency check of the duality, it does provide
interesting evidence for the identification of the
$\SU(2)$-doublet half-hypermultiplets at infinite
coupling as magnetic monopoles of the weak coupling
$\SU(3)$ gauge group.  For, $\SU(3)$ monopoles
which are singlets under the $\SU(6)$ flavor symmetry
also have charge $\pm3$ under the $\U(1)$ flavor
group.  To see this, dress a monopole state
$|M\rangle$ with the fermionic zero-modes of the
6 hypermultiplets in the $\bf3\oplus\bar3$ of the
$\SU(3)$ gauge group.  
This can be done by going to
a point on the Coulomb branch with $v=0$ but $u\neq0$
breaking $\SU(3)\to\U(1)^2$ and leaving one color-component
of each hypermultiplet massless.  These then contribute
6 massless Dirac fermions, giving 12 real zero modes, 
charged under the $\U(1)$ coming from the $\SU(2)\to\U(1)$
breaking in the dual description.  Split these zero modes into
6 creation operators $c_i^\dagger$, and 6 annihilation
operators $c_i$, and take $|M\rangle$ to be the state
annihilated by the $c_i$, so that $|M\rangle$ is an
$\SU(6)$ singlet.  Note that the
$c_i^\dagger$ carry charge $+1$ under the $\U(1)$
flavor symmetry, corresponding to the normalization
we chose above (\ref{weaku1}).  The spectrum of monopole
states is then given by $c_{i_1}^\dagger\cdots c_{i_n}^\dagger
|M\rangle$ with $n=0, \ldots, 6$.  If the $\U(1)$
charge of $|M\rangle$ is $-q$, then these states will
have $\U(1)$ charges $n-q$.  The two $\SU(6)$
singlet states, $|M\rangle$ and $c_1^\dagger\cdots
c_6^\dagger|M\rangle$, therefore have charges $-q$
and $6-q$.  CPT invariance implies these must be
opposite, giving $q=3$.

\paragraph{R-symmetry central charges.}

The central charges for the $\U(1)_R$ and $\SU(2)_R$
factors are proportional because their generators
are both descendants of a single primary, $T$.
It follows from normalizing them on, say, a free multiplet,
as done in appendix D, that $k_{\U(1)_R} = 8 k_{\SU(2)_R}$.

$T$ is the primary of a  
supermultiplet with $R=I=j=\tj=0$ and 
$D=2$.  The $R$-symmetry central charges are proportional to the 
$\vev{TT}$ 2-point function, and their derivative with respect to 
the marginal gauge coupling is proportional to $\vev{\Phi TT}$ where 
$\Phi$ is a chiral supermultiplet with $R=4$, and so vanish by the
same argument as in the last subsection.  Therefore the
$R$-symmetry central charges are also independent of the
gauge coupling, so are the same at weak and infinite coupling.

{}From the $\U(1)_R$ 
free field central charge normalization given in appendix D,
a short calculation gives $k_{\U(1)_R-\SU(3)} = 136/3$ for the weakly
coupled $\SU(3)$ theory.\footnote{Note that this does not
agree with the $N=1$ $\U(1)_R$ central charge, $\t_{RR}$, computed
in \cite{bgisw0507}, because the $N=1$ and $N=2$
$\U(1)_R$ charges are not the same, but are related by
$R_{N=1} = {1\over3} R_{N=2} - {4\over3} I_3$.}
Likewise, the weakly coupled $\SU(2)$ vector multiplet and
two $\bf2$ half-hypermultiplets contribute $k_{\U(1)_R-\SU(2)}=32/3$.
Taking their difference, we deduce that the $E_6$ SCFT contributes
$$
k_{\U(1)_R-E_6} = {104\over 3}.
$$

\section{Infinite coupling in sp(2) w/ $12\cdot\bf4$}

We now quickly run through the same reasoning described in detail 
in the last two sections to support the equivalence
\be\label{sp2=e7}
\Sp(2)\with 12\cdot{\bf 4} =  \SU(2) \with \SCFT_{E_7},
\ee
of the (rank two) Lagrangian SCFT on the left with the strongly 
coupled SCFT on the right.

As reviewed in appendix A.2, the curve for the 
$\Sp(2)$ theory is
\be\label{sp2fac}
y^2 = x\left[ (1-\sqrt f)x^2 -ux-v\right]
\left[ (1+\sqrt f) x^2 -ux -v\right] ,
\ee
where $u$ and $v$ are the Coulomb  branch vevs of 
dimension 2 and 4, respectively, and their associated basis of 
holomorphic one-forms are given in (\ref{basis}).  The infinite 
coupling point is at $f=1$, where the curve degenerates 
to a genus one curve, and it is easy to check that the $\o_u = xdx/y$ 
one-form develops a pair of poles at $x=\infty$, whereas $\o_v = dx/y$ 
remains holomorphic.  Thus, as in the $\SU(3)$ example, when $f=1$,
$v$ is the Coulomb branch vev of a rank 1 SCFT in which $u$ appears 
as a mass deformation, while for $f\sim 1$, $u$ is the vev of an
$\SU(2)$ vector multiplet weakly gauging some global symmetry of 
this rank 1 SCFT.

Isolating the curve of the $\SU(2)$ factor which weakly gauges 
part of the flavor symmetry of the $E_7$ SCFT by setting $v=0$ in 
(\ref{sp2fac}),
and scaling to the non-singular genus one curve by defining 
$\hat y = y/x$, gives
$\hat y^2 = x[(x-u)^2 - f x^2]$ and $\o_u = dx/\hat y$,
which are precisely the curve and one-form \cite{as9509} of the 
scale invariant $\Sp(1)$ $N=2$ superQCD.\footnote{The change of
variables showing its equivalence to the $\SU(2)$ form of 
the curve (\ref{su2curve}) is discussed in \cite{as9509}.}
It is weakly coupled at both $f=0$ and $f=1$ by virtue of the
$\SL(2,\Z)$ S-duality of this theory.

Likewise, isolating the curve of the rank 1 SCFT by setting 
$f=1$, 
and making the change of variables 
\be\label{sp2-sl2c}
x = {(\til v-\tf12 \til u^2)^2\over 
\til x - 2\til u (\til v-\tf12 \til u^2)},\qquad
y = {-(\til v-\tf12 \til u^2)^2\,\,\til y\over
\left(\til x - 2\til u (\til v-\tf12 \til u^2)\right)^2},\qquad
u = 3 \til u ,\qquad
v = \til v - \tf12 \til u^2,
\ee
we get
\be\label{sp2-e7def}
\til y^2 = \til x^3 - \left(2\til v^3 - {3\over2} \til v \til u^4 
+ {1\over2} \til u^6\right) \til x
- \left( 2\til v^4 \til u - 2 \til v^3 \til u^3 
+ {1\over2} \til v \til u^7 - {1\over8} \til u^9 \right),
\qquad \til\o_{\til v} = {d\til x\over\til y} .
\ee
When $\til u=0$, this is the curve of the rank 1 $E_7$ SCFT, reviewed
in appendix B.

With $u\neq0$, this gives a mass deformation of the $E_7$ SCFT curve.
Comparing to the general mass deformation of
the $E_7$ SCFT (\ref{e7-e7def}), gives the $E_7$ adjoint casimirs 
\be\label{sp2-breaking}
M_2 = 2 u , \
M_6 = -2 u^3 , \
M_8 = -\tf32 u^4 , \
M_{10} = 0 , \
M_{12} = \tf12 u^6 , \
M_{14} = \tf12 u^7 , \
M_{18} = -\tf18 u^9 .
\ee
The $M_n$ are given \cite{mn9608} in terms of the 
eigenvalues $m_i$ ($i=1,\ldots,6$) and $m$ of the Cartan subalgebra 
of the $\SO(12)\times\SU(2)$ maximal subgroup of $E_7$ in appendix B.2.
A solution of (\ref{sp2-breaking}) is given by
$m=\sqrt{6\til u}$ and $m_i=0$, showing that the $\til u$ vev
breaks $\SU(2)\times\SO(12) \to \U(1)\times\SO(12)$, thus
identifying the $\SO(12)$ flavor symmetry expected from an
$\Sp(2)$ theory with six massless fundamental 
hypermultiplets.\footnote{As in the footnote in section 2.2, 
some algebra shows that (\ref{sp2-breaking}) is actually consistent 
with three different mass assignments (up to permutations),
$$
\begin{array}{rcrcl}
\mbox{(a):} && m=\sqrt{6\til u},\ \ m_{1,\ldots,6}=0,
&\ \Longleftrightarrow\ &
{\bf su(2)}\times\SO(12) \to 
\ \, {\bf u(1)}\times\SO(12) ,\\
\mbox{(b):} && m=m_{1,\ldots,4}=0,\ \ m_5 = m_6 = \sqrt{3\til u/2},
&\Longleftrightarrow&
{\bf su(2)}\times\SO(12) \to 
{\bf su(2)}\times\SO(8)\times\SU(2)\times\U(1) ,\\
\mbox{(c):} && m=\sqrt{3\til u/2}, m_{1,2,3,4,5}=-m_6=\sqrt{3\til u/8},
&\Longleftrightarrow&
{\bf su(2)}\times\SO(12) \to 
\ \, {\bf u(1)}\times\SU(6)\times\SU(2) ,
\end{array}
$$
where the corresponding adjoint breaking patterns
of the $\SU(2)\times\SO(12)$ maximal subgroup of $E_7$ are shown
on the right.
The (a) breaking, which manifestly leaves an unbroken $\SO(12)$ factor,
is the one described above.
The (b) breaking pattern actually
also leaves an $\SO(12)$ unbroken.  This is not
manifest because the unbroken $\SO(12)$
does not coincide with the $\SO(12)$ used for the basis of Casimirs.
The (c) breaking pattern is different, corresponding to adjoint
breaking of the $\SU(3)$ factor in the $\SU(3)\times\SU(6)\subset E_7$
maximal subgroup, and does not give the expected global
symmetry group of the original $\Sp(2)$ SCFT.}

The vanishing of the beta function for the $\SU(2)$ gauge
coupling implies, as argued in section 3, that the
central charge of the SCFT flavor current algebra be
given by (\ref{CFTcentral}).
Applying this to the present example using group theory 
data from appendix C, we compute 
the central charge of the $E_7$ SCFT current algebra to be
\be
k_{E_7} = 2 \cdot T({\bf 3}) / I_{\SU(2)\hra E_7} = 8.
\ee
This can be independently checked by comparing the
$\SO(12)$ flavor algebra central charge in the weak
coupling and infinite coupling descriptions.
At weak coupling the
half-hypermultiplets transform as a $\bf(4,12)$ under 
the $\Sp(2)\times\SO(12)$ combined
gauge and flavor symmetries.  Thus the central charge of 
the $\SO(12)$ flavor current is by (\ref{khypers})
\be\label{weakso12}
k_{\SO(12){\rm -weak}} = 4\cdot T({\bf12}) = 8.
\ee
At infinite coupling it follows by (\ref{kGform2}) that
\be\label{strongso12}
k_{\SO(12){\rm -strong}} = I_{\SO(12)\hra E_7} \, k_{E_7} = 8.
\ee
The agreement of (\ref{weakso12}) and (\ref{strongso12}) is
a non-trivial check of the duality.

Finally, by comparing $\U(1)_R$ central charges in the
two dual theories, we deduce that the $E_7$ SCFT contributes
$$
k_{\U(1)_R-E_7} = k_{\U(1)_R-\Sp(2)} -
k_{\U(1)_R-\SU(2)} = {176\over3}-8={152\over3}.
$$

\acknowledgments It is a pleasure to thank A. Buchel, K. Intriligator,
R. Plesser, A. Shapere, M. Strassler, E. Witten, and J. Wittig for 
interesting and helpful discussions on infinite coupling singularities, 
some over many years.  
PCA is supported in part by DOE grant DOE-FG02-84-ER40153. 
NS is supported by in part by DOE grant DOE-FG02-90-ER40542.

\appendix
\section{Rank 2 Lagrangian SCFTs}

Here we briefly review the systematics of Lagrangians for $N=2$
superQCD, and then collect from the literature the known 
curves for the rank 2 scale-invariant superQCDs.  

\subsection{Scale-invariant rank 2 superQCDs.}

It is a straightforward group theory exercise to determine all
the rank two $N=2$ theories with vanishing beta function.
They are listed in table \ref{tab.rank2}, along with their flavor 
symmetries, and whether or not they have infinite coupling points 
in their spaces of couplings.

\TABLE[ht]
{\begin{tabular}{||l||l|l|c|c||} \hline
\# & Gauge group & Half-hypermultiplets & Flavor symmetry & Infinite coupling?\\
\hline\hline
1  & $\SU(2)\times\SU(2)$ & $4\cdot{\bf(2,2)}$ & $\Sp(2)$ & no? \\ \hline
2  & $\SU(2)\times\SU(2)$ & $2\cdot({\bf(2,2)\oplus(2,1)\oplus(1,2)})$ 
& $\Sp(1)\times\SO(2)^2$ & no \\ \hline
3  & $\SU(3)$ & $2\cdot{\bf8}$ & $\Sp(1)$ & no \\ \hline
4  & $\SU(3)$ & $6\cdot({\bf3\oplus\bar3})$ & $\U(6)$ & yes \\ \hline
5  & $\SU(3)$ & $\bf3\oplus\bar3\oplus6\oplus\bar6$ 
& $\U(1)^2$ & yes \\ \hline
6  & $\Sp(2)$ & $2\cdot{\bf10}$ & $\Sp(1)$ & no \\ \hline
7  & $\Sp(2)$ & $12\cdot{\bf4}$ & $\SO(12)$ & yes \\ \hline
8  & $\Sp(2)$ & $8\cdot{\bf4}\oplus2\cdot{\bf5}$ & $\SO(8)\times\Sp(1)$ & 
no \\ \hline
9  & $\Sp(2)$ & $4\cdot{\bf4}\oplus4\cdot{\bf5}$ & $\SO(4)\times\Sp(2)$ & 
? \\ \hline
10 & $\Sp(2)$ & $6\cdot{\bf5}$ & $\Sp(3)$ & yes \\ \hline
11 & $G_2$ & $2\cdot{\bf14}$ & $\Sp(1)$ & no \\ \hline
12 & $G_2$ & $8\cdot{\bf7}$ & $\Sp(4)$ & yes \\ \hline
\end{tabular}
\caption{Rank 2 scale-invariant $N=2$ gauge theories which are not
products of two rank 1 theories. \label{tab.rank2}}}

The table does not include the three theories which are products
of two decoupled rank 1 scale-invariant gauge theories.  These
rank 1 theories are the $\SU(2)$ theories with either two adjoint 
or eight fundamental half-hypermultiplets.  The first is the $N=4$
$\SU(2)$ theory, and both were examined in \cite{sw9408}.  Both
have an $\SL(2,\Z)$ S-duality group, and so only have weakly
coupled limits.

Theories \#1 and \#2 in table \ref{tab.rank2} have two marginal
couplings.  Their low energy effective actions were found in
\cite{w9703}.  The S-duality group of theory \#2 was determined
in \cite{a9706,ab9910}, and the self-duality of the $\SU(2)$ factors
were found to eliminate any infinite coupling limits.  A similar
analysis has not been performed for theory \#1, but it seems probable
that it has no infinite coupling limits, for the same reason.  Indeed,
taking the coupling of one of the $\SU(2)$ factors small, the
$\SL(2,\Z)$ duality of the other factor eliminates any infinite-coupling
limit in its coupling.  Continuing this to strong coupling in the
first factor then disallows any one-(complex-)dimensional submanifolds
of infinite coupling.  However, the possibility remains of an isolated
infinite coupling point at strong coupling in both factors.

Theories \#3, \#6, and \#11 are $N=4$ theories, all of which are
self-dual \cite{OMduality}.  (Note that $\Sp(2)\simeq\SO(5)$, so that
the $\Sp(2)$ theory is actually self-dual.)  

The low energy effective action of theory \#8 was determined in
\cite{asty9611,dls9612} and indicates an $\SL(2,\Z)$ duality group
with only weakly coupled limits.  
The low energy effective action of \#9, and whether it has an
infinite coupling limit, is not known.

The effective actions of the remaining theories imply that
they have infinite coupling limits, as shown for \#4 in \cite{aps9505},
\#5 in \cite{lll9805}, \#7 and \#10 in \cite{as9509}, and
\#12 in \cite{acsw0504}.  We now describe their effective actions
in more detail.

\subsection{Rank 2 effective actions}

The effective actions on the Coulomb branches of the above-mentioned 
theories are most conveniently encoded in the associated genus 2
curve.  Recall \cite{sw9408} (see, \eg, \cite{acsw0504} for a brief 
review) that the low energy effective action on the Coulomb branch 
of a theory with rank $r$ gauge group $G$ is an $N=2$ $\U(1)^r$ theory, 
parametrized by the $r$ vevs of the complex scalars in each $\U(1)$ 
vector multiplet.  Taking $\{u^k\}$ as $r$ good complex coordinates
on the Coulomb branch, the matrix $\t_{ij}(u^k)$ of complex 
$\U(1)^r$ couplings and the central charge $Z(u^k)$ of the $N=2$ 
algebra can be encoded (at least for $r<4$) by a holomorphic family 
of genus $r$ Riemann surfaces $\Sigma(u^k)$ together with a specified 
basis $\{\o_i\}$ of the $r$ holomorphic one-forms on $\Sigma$:
\be\label{swmap}
\t_{ij} = A_{ik} (B^{-1})^k_{\ j}, 
\qquad
{\del Z(\g) \over\del u^k} = \oint_\g \o_k ,
\ee
where $A_{ik} := \oint_{\a_i}\o_k$, $B^{\ j}_k := \oint_{\b^j}\o_k$, 
$\{\a_i,\b^j\}$ are a basis of homology one forms with
canonical intersection matrix, and the homology class of the 
contour $\g$ is determined by the electric and magnetic charges
of the state.  Note in particular, that under holomorphic
changes of variables on the Coulomb branch $u^k \to \til u^k$,
the holomorphic one forms transform as $\o_k \to \til\o_k = 
(\del u^\ell/\del\til u^k) \o_\ell$, since it is the central
charge that remains invariant.

Turning on masses $m_a$ in these theories corresponds to deformations
of the Riemann surfaces $\Sigma(u^k, m_a)$ such that there
exists a central charge $Z$ which depends linearly on the masses
with integer coefficients \cite{sw9408}.  This means that 
the second equation in (\ref{swmap}) can be integrated  to
$Z(\g) = \oint_\g \l$, where $\l$ is a meromorphic one-form 
whose residues are integral linear combinations of the $m_a$, and
which satisfies 
\be\label{swform}
{\del\l\over\del u^k} = \o_k + df_k,
\ee
where $df_k$ are total derivatives on the curve $\Sigma$.

Specializing to rank 2, call the two coordinates on the Coulomb
branch $u$ and $v$.  Since all genus 2 Riemann surfaces are hyperelliptic,
they can all be described as complex curves in a 2-dimensional
projective space of the form $y^2={\cal P}(x)$ where $\cal P$
is a fifth- or sixth-order polynomial in $x$.  This realizes the
Riemann surface as a 2-sheeted cover of the complex $x$-plane
(plus the point at infinity) branched at six points (the zeros
of $\cal P$).  $\cal P$ can vary holomorphically with $u$ and $v$, and
degenerations of the curve correspond to collisions of the branch points.
It is possible, by suitable coordinate changes, to choose the basis of 
holomorphic one forms to be
\be\label{basis}
\o_u = {xdx\over y}, \qquad
\o_v = {dx\over y}.
\ee
We will use such coordinates in what follows.

\paragraph{su(3) w/ $6\cdot\bf(3\oplus\bar3)$ and
su(3) w/ $\bf3\oplus\bar3\oplus6\oplus\bar6$.}
The curve for the scale-invariant theory with
6 fundamental hypermultiplets is \cite{aps9505,as9509}
\be\label{A2+6.3}
y^2 = (x^3 - u x -v)^2 - f x^6
\ee
where $f$ is a holomorphic function of the microscopic gauge coupling
$\tau$.  At weak coupling $f\sim e^{2\pi i \tau}$.  The curve
degenerates whenever the 
discriminant in $x$ of its right side vanishes.  For (\ref{A2+6.3})
the discriminant is $f^3(f-1)$ times factors that depend on $u$
and $v$ moduli.  The vanishing of the moduli-dependent factors 
determines submanifolds on the Coulomb branch where various
dyons become massless.  The coupling-dependent prefactor, on the
other hand, indicates values of the couplings where there are
singularities in the effective action everywhere on the
Coulomb branch:  the curve becomes singular at $f=0$ and $f=1$,
irrespective of the values of the Coulomb branch vevs $u$, $v$.
The $f=0$ singularity has the interpretation as the weak coupling 
limit of the $\SU(3)$ gauge theory, while the $f=1$ singularity
is the infinite coupling singularity which is the subject of
this paper.

The curve for the scale-invariant theory with one symmetric and
one antisymmetric hypermultiplet is also given by (\ref{A2+6.3}).
This can be deduced from \cite{lll9805}.  
These are nevertheless different theories.  In particular they
have different global symmetry groups and different mass 
deformations.  This gives an example of distinct
scale invariant theories with the same scale-invariant
form, but different mass deformations.

The 6-flavor theory has a $\U(1)\times\SU(6)$ global flavor symmetry
group, and therefore a deformation by 6 mass parameters: the 
$\U(1)$ mass $M$ of dimension 1, and the five adjoint Casimirs
of $\SU(6)$, $S_n$, with dimension $n$, for $n=2,\ldots,6$.  
The explicit mass deformations of (\ref{A2+6.3}) for the 6 
flavor theory is given in \cite{aps9505,as9509}:
\be\label{A2+6.3massive}
y^2 = \left[(x+\sqrt{1-f} M)^3 - u (x+\sqrt{1-f} M) - v\right]^2 
- f \left[ x^6 - S_2 x^4 - S_3 x^3 - S_4 x^2 - S_5 x -S_6 \right] .
\ee
If the $m_i$ are the eigenvalues of the $\SU(6)$ adjoint mass matrix
satisfying $\sum_i m_i=0$, then the $S_n$ Casimirs are given by 
$S_n := \sum_{i_1<\cdots<i_n} m_{i_1} \cdots m_{i_n}$. 
With this specific mass dependence of the curve, one can
then integrate (\ref{swform}) to find $\l$ and thus a
central charge $Z(u,v;M,m_i)$ which depends linearly on the
mass eigenvalues with integer coefficients.  It is worth
emphasizing that the normalization of the masses and
the specific basis of the flavor symmetry adjoint Casimirs,
$S_n$, that enter into (\ref{A2+6.3massive}) are determined by 
linear integral dependence of the central charge on the
masses.

The symmetric plus antisymmetric
theory has global flavor symmetry $\U(1)\times\U(1)$ and
two mass deformation parameters both of dimension 1.
The mass deformation of (\ref{A2+6.3}) for the symmetric plus 
antisymmetric theory is not known, though the deformation of the 
double-cover curve is given in \cite{lll9805}.

\paragraph{sp(2) w/ $12\cdot\bf4$ and  sp(2) w/ $6\cdot\bf5$.}
The scale-invariant curve is, in either case, \cite{as9509}
\be\label{C2+6.4}
y^2 = x(x^2 - u x -v)^2 - f x^5 ,
\ee
and degenerates for all $u$ and $v$ whenever the coupling
$f=0$ or $f=1$.  The $f=0$ singularity is the weak coupling
limit of the $\Sp(2)$ gauge theory, while the $f=1$ singularity
is the new infinite coupling limit. $u$ and $v$ are Coulomb
branch vevs of dimension 2 and 4, respectively.
This is another example of the same scale-invariant curve having
two inequivalent mass deformations.  

The theory with 12 half-hypermultiplets
in the 4-dimensional representation has global flavor symmetry group
$\SO(12)$, and therefore 6 mass parameters with dimensions
2, 4, 6, 6, 8, and 10.  This mass deformations of (\ref{C2+6.4}) is
\cite{as9509}
\be
y^2 = x(x-u)^2 - 2 \sqrt f (x-u) s_6
- f (x^5 - S_2 x^4 + S_4 x^3 - S_6 x^2 + S_8 x - S_{10}).
\ee
If $\pm m_i$, $i=1,\ldots,6$ are the eigenvalues of the $\SO(12)$
adjoint mass matrix, then the mass parameters appearing in the
curve are the Casimirs $S_{2n} :=
\sum_{i_1<\cdots<i_n} m_{i_1}^2\cdots m_{i_n}^2$, and $s_6 :=
\prod_i m_i$. 

The theory with 6 half-hypermultiplets in
the 5-dimensional representation has global flavor symmetry
group $\Sp(3)$ and 3 mass parameters of dimensions 2, 4, and 6.
This mass deformation of (\ref{C2+6.4}) is also given in \cite{as9509}.

\paragraph{G$_{\bf2}$ w/ $8\cdot\bf7$.}
The curve for the scale invariant theory is \cite{acsw0504}
\be\label{G2+8.7}
v y^2 = (x^3 - uv x - 2v^2)^2 - f x^6 ,
\ee
and again has a weak coupling singularity at $f=0$ and
an infinite coupling singularity at $f=1$.  $u$ and $v$
are Coulomb branch vevs of dimension 2 and 6.  The
global flavor symmetry of this theory is $\Sp(4)$, and
so the curve should have a 4-parameter mass deformation
with masses of dimension 2, 4, 6, and 8; however the
explicit form of this deformation is not known.

\section{The E$_{\bf6}$ and E$_{\bf7}$ rank 1 SCFTs}

The curves and one-forms encoding the effective action on the
Coulomb branch for the rank 1 $N=2$ SCFTs with 
$E_n$ global symmetry groups was first worked out in
\cite{mn9608}.  Since these are all rank one theories, the
curves are elliptic (genus 1 Riemann surfaces) of the
form $\til y^2 = \til x^3 + \ldots$, and we choose
the basis of holomorphic one forms to be
$\o = d\til x/\til y$.

\subsection{E$\bf{_6}$}

The curve for the scale-invariant $E_6$ SCFT (\ie, without
mass deformations) is
\be\label{e6crv}
\til y^2 = \til x^3 - \til v^4.
\ee
The Coulomb branch vev, $\til v$, has mass dimension 3.

The maximal mass deformation of this curve is
\be\label{e6-e6def}
\til y^2 = \til x^3 - (M_2 \til v^2+ M_5 \til v + M_8) \til x 
- ( \til v^4 + M_6 \til v^2 + M_9 \til v + M_{12}).
\ee
Here we have added all possible terms which deform the
complex structure of the curve.  Terms proportional to
$\til x^2$ do not appear since they can be reabsorbed in a shift
in the $\til x$ variable.  Likewise, an $M_3 \til v^3$ term does not
appear since its deformation is simply a shift in the
Coulomb branch vev $\til v$.  The subscripts of the remaining
six mass parameters record their mass dimensions.  They
are the dimensions of the adjoint Casimirs of $E_6$, hinting
that they break an $E_6$ global symmetry group.

To confirm this, one must construct a central charge $Z$
which depends linearly on the $E_6$ mass eigenvalues
with integer coefficients \cite{sw9408}.  
Calling the six mass eigenvalues of the $E_6$ mass matrix
$m_a$, $a=1,\ldots,6$, then the integers $n^a=\del Z(\gamma)/\del m_a$
are the ``quark number" charges of the generically unbroken
$\U(1)^6$ flavor symmetry.  Thus one must find a specific
$E_6$-Weyl-invariant polynomial form for the Casimirs in
terms of the eigenvalues, $M_n(m_a)$, such that 
the second equation in (\ref{swmap}) can be integrated  to
$Z(\g) = \oint_\g \l$, where $\l$ is a meromorphic one-form 
whose residues are integral linear combinations of the $m_a$, and
satisfying $\del\l/\del \til u = \o + df$ where $df$ is a total
derivative on the curve.   Following a method described in
section 17 of \cite{sw9408}, this was done for the
$E_n$ curves in \cite{mn9608,nty9903}.

The result of \cite{mn9608} for the $E_6$ mass deformation is
\bea\label{e6crv-e6}
M_2 &=& -\tf{1}{3} P_2, \qquad\qquad
M_5 \ =\ \tf{2}{3} P_5, \qquad\qquad
M_6 \ =\ \tf{2}{3} P_6 - \tf{7}{108} P_2^3 
\nonumber\\
M_8 &=& -\tf{7}{432} P_2^4 + \tf{11}{45} P_2P_6 - \tf{8}{15} P_8,
\qquad\qquad\quad
M_9 \ =\ \tf{1}{18} P_2^2P_5 - \tf{8}{21} P_9
\\
M_{12} &=& \tf{32}{135} P_{12} - \tf{298}{18225}P_2^2P_8
-\tf{101}{218700} P_2^3P_6 + \tf{13}{405} P_6^2 
-\tf{49}{1049760} P_2^6 - \tf{19}{3645} P_2P_5^2,
\nonumber
\eea
where the $P_n$ are the basis of $E_6$ Casimirs given by
\cite{lw90}.  These Casimirs can be written in
terms of the Casimirs $\til S_n$ and $\til T$ of the $\SU(6)\times\SU(2)$ 
maximal subgroup of $E_6$ as\footnote{In \cite{lw90,mn9608} 
the $E_6$ Casimirs were expressed in terms of a basis of Casimirs of
the $\SO(10)\times\U(1)$ maximal subgroup of $E_6$.  For our purposes
it is much more convenient to express them in terms of a basis of
$\SU(6)\times\SU(2)$ Casimirs.  To do that, we followed the
method of \cite{lw90} by computing the character of
the $\bf 27$ of $E_6$ using the fact that under $\SU(6)\times\SU(2)
\subset E_6$, the $\bf 27$ decomposes as $\bf 27 = (6,2)\oplus
(\overline{15}, 1)$.}
\bea\label{e6-casimirs}
P_2&=& -6\, \til T + 6\, \til S_2 ,
\qquad\qquad\qquad\qquad
P_5\ =\ 12\, \til T \til S_3 + 12\, \til S_5 ,
\nonumber\\
P_6&=& -20\, \til T^3 + 64\, \til T^2 \til S_2 
- 64\, \til T \til S_2^2 + 20\, \til S_2^3 
- 3\, \til S_3^2 + 20\, \til T \til S_4 + 4\, \til S_2 \til S_4 
- 24\, \til S_6 ,
\nonumber\\
P_8&=& 15\, \til T^4 - 76\, \til T^3 \til S_2 + 112\, \til T^2 \til S_2^2 
- 76\, \til T \til S_2^3 + 15\, \til S_2^4 - 3\, \til T \til S_3^2 
- 12\, \til S_2 \til S_3^2 - 50\, \til T^2 \til S_4 
\nonumber\\ &&\ \mbox{}
+ 54\, \til T \til S_2 \til S_4 + 16\, \til S_2^2 \til S_4 
- 10\, \til S_4^2 + 15\, \til S_3 \til S_5 
+ 186\, \til T \til S_6 - 66\, \til S_2 \til S_6 ,
\nonumber\\
P_9&=& 56\, \til T^3 \til S_3 - 140\, \til T^2 \til S_2 \til S_3 
+ 56\, \til T \til S_2^2 \til S_3 - 56\, \til T \til S_3 \til S_4 
+ 140\, \til T^2 \til S_5 - 56\, \til T \til S_2 \til S_5 
\nonumber\\ &&\ \mbox{}
+ 56\, \til S_2^2 \til S_5 + 28\, \til S_4 \til S_5 
- 84\, \til S_3 \til S_6 ,
\\
P_{12}&=& \til T^6 - 22\, \til T^5 \til S_2 + 67\, \til T^4 \til S_2^2 
- 72\, \til T^3 \til S_2^3 + 67\, \til T^2 \til S_2^4 
- 22\, \til T \til S_2^5 + \til S_2^6 - 33\, \til T^3 \til S_3^2 
\nonumber\\ &&\ \mbox{}
+ 72\, \til T^2 \til S_2 \til S_3^2 + 18\, \til T \til S_2^2 \til S_3^2 
- 12\, \til S_2^3 \til S_3^2 + 3\, \til S_3^4 
- 28\, \til T^4 \til S_4 + 66\, \til T^3 \til S_2 \til S_4 
\nonumber\\ &&\ \mbox{}
- 118\, \til T^2 \til S_2^2 \til S_4 + 4\, \til T \til S_2^3 \til S_4 
+ 16\, \til S_2^4 \til S_4 + 50\, \til T \til S_3^2 \til S_4 
- 8\, \til S_2 \til S_3^2 \til S_4 - 64\, \til T^2 \til S_4^2 
\nonumber\\ &&\ \mbox{}
+ 58\, \til T \til S_2 \til S_4^2 
- 12\, \til S_2^2 \til S_4^2 - 20\, \til S_4^3 
- 147\, \til T^2 \til S_3 \til S_5 
- 194\, \til T \til S_2 \til S_3 \til S_5 
+ 26\, \til S_2^2 \til S_3 \til S_5 
\nonumber\\ &&\ \mbox{}
+ 45\, \til S_3 \til S_4 \til S_5 + 251\, \til T \til S_5^2 
+ 19\, \til S_2 \til S_5^2 + 522\, \til T^3 \til S_6 
- 342\, \til T^2 \til S_2 \til S_6 + 584\, \til T \til S_2^2 \til S_6 
\nonumber\\ &&\ \mbox{}
- 44\, \til S_2^3 \til S_6 
- 87\, \til S_3^2 \til S_6 - 590\, \til T \til S_4 \til S_6 
+ 26\, \til S_2 \til S_4 \til S_6 - 78\, \til S_6^2 .
\nonumber 
\eea
We have defined the $\til S_n$ and $\til T$ Casimirs by
\be\label{su6su2-casimirs}
\til S_n := \!\!\sum_{i_1<\cdots<i_n}\!\! m_{i_1}\cdots m_{i_n} 
\qquad  (n=2,\dots,6), \qquad\qquad
\til T := m^2 ,
\ee
where $\pm m$ are the mass eigenvalues in the $\SU(2)$ factor, 
and $m_i$ for $i=1,\ldots,6$ with $\sum_i m_i=0$ are the mass 
eigenvalues for the Cartan subalgebra of the $\SU(6)$ factor.
The associated meromorphic one-form is given in \cite{mn9608}, but
will not be needed here.

\subsection{E$\bf_7$}

The curve for the scale-invariant $E_7$ SCFT is
\be\label{e7crv}
y^2 = x^3 - 2 u^3 x.
\ee
The Coulomb branch vev $u$ has mass dimension 4.
The factor of 2 is a normalization (of $u$) chosen to
match to that of \cite{mn9608}.
The maximal mass deformation of this curve is
\be\label{e7-e7def}
y^2 = x^3 - (2u^3 + M_8 u+ M_{12}) x 
- (M_2 u^4 + M_6 u^3 + M_{10} u^2 + M_{14} u + M_{18}).
\ee
The deformation parameters $M_n$ are adjoint Casimirs of 
$E_7$ determined in \cite{mn9608} to be the following 
expressions\footnote{We have shifted our Coulomb vev $u$
relative to that of \cite{mn9608} to eliminate a $u^2 x$
term, and consequently also shifted their $P_n$ Casimirs
to the $M_n$ values shown in (\ref{e7shift}).}
in terms of the Casimirs $T_n$, $t_6$, and $U$ 
of the $\SO(12)\times\SU(2)$ maximal subgroup of $E_7$:
\bea\label{e7shift}
M_2 &=& P_2, 
\quad
M_6 \ =\  P_6 - \tf23 P_2 P_4,
\quad
M_8 \ =\ P_8 - \tf16 P_4^2,
\quad
M_{10} \ =\ P_{10} - \tf12 P_4 P_6 + \tf16 P_2 P_4^2,
\nonumber\\ 
M_{12} &=& P_{12} - \tf16 P_4 P_8 + \tf1{54} P_4^3,
\qquad\quad
M_{14} \ =\ P_{14} - \tf13 P_4 P_{10} + \tf1{12} P_4^2 P_6
- \tf1{54} P_2 P_4^3,
\nonumber\\
M_{18} &=& P_{18} - \tf16 P_4 P_{14} + \tf1{36} P_4^2 P_{10}
- \tf1{216} P_4^3 P_6 + \tf1{1296} P_2 P_4^4,
\eea
where
\bea\label{e7-casimirs}
P_2 &=&
U
+\tf23 \tT,
\qquad\qquad\qquad\qquad\qquad\qquad\ 
P_4 \ =\
\tf1{12} \tT^2
+T_4,
\nonumber\\
P_6 &=&
-\tf1{108} \tT^3
+\tf13 \tT T_4
+\tf{20}3 t_6
+\tf23 T_6.
\qquad\quad
P_8 \ =\ 
8 U t_6
+\tf23\tT t_6
+\tf23 \tT T_6
-2 T_8,
\nonumber\\
P_{10} &=&
4 T_{10}
-\tf{10}3 U \tT t_6
-\tf{29}{18} \tT^2 t_6
+\tf{22}3 T_4 t_6
+\tf5{36} \tT^2 T_6
+\tf13 T_4 T_6
-2 U T_8
-\tf23 \tT T_8,
\nonumber\\
P_{12} &=&
4 U T_{10}
-\tf14 {\tT^3} t_6
+\tT T_4 t_6
+\tf43 {t_6^2}
-\tf43 t_6 T_6
+\tf13 {T_6^2}
+\tf14 \tT^2 T_8-T_4 T_8,
\\
P_{14} &=&
-\tf83 U \tT T_{10}
-\tT^2 T_{10}
+4 T_{10} T_4
+\tf1{24} \tT^4 t_6
-\tf23 \tT^2 T_4 t_6
+2 T_4^2 t_6
+\tf{32}3 U t_6^2 
-\tf49 \tT t_6^2
\nonumber\\ &&{}
-\tf{16}3 U t_6 T_6
-\tf29 \tT t_6 T_6
+\tf29 \tT T_6^2
+\tf1{12} \tT^3 T_8
-\tf13 \tT T_4 T_8
+\tf43 t_6 T_8
-\tf23 T_6 T_8,
\nonumber\\
P_{18} &=&
\tf1{16} \tT^4 T_{10}
-\tf12 \tT^2 T_{10} T_4
+T_{10} T_4^2 
+\tf{16}3 U T_{10} t_6
+ U \tT^2 t_6^2
+\tf16 \tT^3 t_6^2
-\tf23 \tT T_4 t_6^2
-\tf{16}{27} t_6^3
\nonumber\\ &&{}
-\tf83 U T_{10} T_6
-\tf1{12} \tT^3 t_6 T_6
+\tf13 \tT T_4 t_6 T_6
+\tf89 t_6^2 T_6
-\tf49 t_6 T_6^2
+\tf2{27} T_6^3
- 2U \tT t_6 T_8
\nonumber\\ &&{}
-\tf16 \tT^2 t_6 T_8
+\tf23 T_4 t_6 T_8
+\tf1{12} \tT^2 T_6 T_8
-\tf13 T_4 T_6 T_8
+ U T_8^2 .
\nonumber
\eea
Here $\tT := T_2 - U$ 
and the $T_n$, $t_6$, and $U$ Casimirs are defined by
\be\label{so12su2-casimirs}
T_{2n} := \!\!\sum_{i_1<\cdots<i_n}\!\! m_{i_1}^2\cdots m_{i_n}^2 
\quad (n=1,\dots,5), \qquad
t_6 := \prod_i m_i ,
\qquad
U := \ \ m^2 ,
\ee
where $\pm m$ are the mass eigenvalues in the $\SU(2)$ factor, 
and $\pm m_i$ for $i=1,\ldots,6$ are the mass 
eigenvalues for the Cartan subalgebra of the $\SO(12)$ factor.

\section{Lie algebra indices}

We normalize the inner product on the root space of simple Lie 
algebras by choosing the long roots to have length $\sqrt2$.
This is the normalization in which the quadratic index of the
$\bf n$ of $\SU(n)$ is $T({\bf n})=1$.

If under an embedding $G\subset H$ of Lie algebras, the
generators $\{\t^\a\}$ of $G$ are related to the 
generators $\{t^a\}$ of $H$ by $\t^\a = \sum_a c^\a_a t^a$,
and a representation
$\br$ of $H$ decomposes as $\oplus_i \br_i$ under $G$, then
the Dynkin index of embedding of a Lie algebra $G$ in $H$ is
\be\label{dynkinindex}
I_{G\hra H} 
:= {\sum_{\a,a} (c^\a_a)^2 \over{\rm dim}(G)} 
= {\sum_i T(\br_i) \over T(\br)} ,
\ee
independent of the choice of $\br$.

The examples in the body of the paper all turn out
to give Dynkin index 1:  From tables, \eg\ \cite{patera},
the $\bf 27$ of $E_6$ decomposes as ${\bf27} = ({\bf2},{\bf6}) \oplus
({\bf1},{\bf15})$ under the maximal subalgebra $E_6\supset
\SU(2)\times\SU(6)$.  Thus for these embeddings
\be
I_{\SU(2)\hra E_6} =
{6 \cdot T({\bf2}) + 15 \cdot T({\bf1}) \over T({\bf27}) } 
=1,
\qquad
I_{\SU(6)\hra E_6} = 
{2\cdot T({\bf 6})+1\cdot T({\bf 15}) \over T({\bf 27}) }
= 1.
\ee
Similarly, $\bf56=(2,12)\oplus(1,32)$ under
$E_7 \supset \SU(2)\times\SO(12)$, giving
\be
I_{\SU(2)\hra E_7} =
{12 \cdot T({\bf2}) + 32 \cdot T({\bf1}) \over T({\bf56}) } 
=1,
\qquad
I_{\SO(12)\hra E_7} =
{2\cdot T({\bf 12})+1\cdot T({\bf 32}) \over T({\bf 56}) }
= 1.
\ee

\section{Normalization of central charges}

{}From \cite{bgisw0507}, the $\U(1)_i$--$\U(1)_j$ current 
2-point function for free fields is given by
\be\label{u1u1norm}
\vev{J_\m^i(x)J_\n^j(0)} = 
\left(\sum_b q^i_b q^j_b + 2\sum_f q^i_f q^j_f\right) 
{1\over 4\pi^4}
{x^2 g_{\m\n} - 2 x_\m x_\n \over x^8}
\ee
where $b$ runs over complex scalars and $f$ over Weyl fermions,
and $q^i_{b,f}$ are their charges under the $\U(1)_i$ group, namely,
$[Q^i,\phi_b]=q^i_b \phi_b$ and $[Q^i,\psi_f]=q^i_f \psi_f$.  The
$\U(1)_i$ charges are related to the currents in the usual way by
$Q^i = \int  d^3x J_0^i$.

For $n$ half-hypermultiplets of charges $q^i_h=\d^i_h$, $h=1,\ldots,n$,
so that $Q^i$ counts the number of $i$th half-hypermultiplets,
this then gives
\be\label{U1U1fnc}
\vev{J_\m^i(x)J_\n^j(0)} = 
{3\d^{ij}\over 4\pi^4} {x^2 g_{\m\n} - 2 x_\m x_\n \over x^8} .
\ee
These $n$ $\U(1)$'s form the Cartan subalgebra of the
$\U(n)$ flavor symmetry rotating the $n$ half-hypermultiplets.  
A basis of the $n^2$ Hermitian $\U(n)$ generators in the 
fundamental are $Q^{(ij)}$ and $Q^{[ij]}$ with matrix
elements $[Q^{(ij)}]_{k\ell} = {1\over2}(\d^i_k\d^j_\ell+\d^i_\ell
\d^j_k)$ and $[Q^{[ij]}]_{k\ell} = {i\over2}(\d^i_k\d^j_\ell-
\d^i_\ell \d^j_k)$.  Thus the $\U(1)_i$ generator $Q^i=Q^{(ii)}$
(including normalization).  Furthermore, it is easy to check that 
this is an orthogonal basis:
\be\label{uNbasis}
\tr(Q^{(ij)}Q^{(k\ell)})=\d^{(ij)(k\ell)},\qquad
\tr(Q^{(ij)}Q^{[k\ell]})=0,\qquad
\tr(Q^{[ij]}Q^{[k\ell]})=\d^{[ij][k\ell]},
\ee
where the symmetrized delta symbols are defined as
$\d^{(ij)(k\ell)} = {1\over2}(\d^{ik}\d^{j\ell}+\d^{i\ell}
\d^{jk})$ and $\d^{[ij][k\ell]} = {1\over2}(\d^{ik}\d^{j\ell}
-\d^{i\ell}\d^{jk})$.  The traces (\ref{uNbasis}) also show
that these generators are normalized as in appendix C.
Thus, comparing (\ref{flavorOPE}) and (\ref{U1U1fnc}) gives
the central charge
\be\label{kfree}
k_{\rm free} = 1
\ee
of the $\U(n)$ flavor symmetry of $n$ free half-hypermultiplets.

Say we have a global symmetry $H$ with conserved currents $J^A_\m$
and the current-current OPE (\ref{flavorOPE}) has central charge 
$k_H$.  Consider a simple subalgebra $G\subset H$.
Similar considerations show that the central 
charge of the $G$ currents, $J^a_\m$, are calculated by
\be\label{kGform2}
k_G = k_H I_{G\hra H}.
\ee
 
We now compute the contribution of $n$ half-hypermultiplets
to the central charge of a weakly gauged subgroup $G\subset\U(n)$
of the $\U(n)$ (free) flavor symmetry.  Suppose $G$ is such
that the $n$ half-hypermultiplets, which form the $\bf n$ of
$\U(n)$, transform as ${\bf n}=\oplus_i \br_i$ under $G$.  
(\ref{kGform2}), (\ref{kfree}), and (\ref{dynkinindex}) then imply
\be\label{khypers}
k_G = k_{\rm free} \, I_{G\hra\U(n)}
= {1\over T({\bf n})} \sum_i T(\br_i) 
= \sum_i T(\br_i) .
\ee
This is the expected result: the contribution of half-hypers
to the beta function is proportional to the sum of the indices
of the representations of the half-hypers.

Finally, we compute the central charges of the $\U(1)_R\times
\SU(2)_R$ symmetry for
free half-hypermultiplets and vector multiplets.
A free half-hypermultiplet in a scale-invariant $N=2$ gauge theory 
has a complex scalar with $R=0$ and $(I,I_3)=(1/2,1/2)$ and a 
Weyl fermion with $R=1$ and $I=0$, which, 
together with (\ref{u1u1norm}) and (\ref{flavorOPE}), gives 
$k_{\U(1)_R-{\rm half-hyper}}=2/3$.  A free vector multiplet has 
an $R=2$, $I=0$ complex scalar, a doublet of $R=1$, $I=1/2$ Weyl 
fermions, and an $R=0$, $I=0$ vector field, giving 
$k_{\U(1)_R-{\rm vector}}=8/3$.
Similarly, $k_{\SU(2)_R-{\rm half-hyper}}=1/12$, and
$k_{\SU(2)_R-{\rm vector}}=1/3$.


\begin{thebibliography}{99} 

\bibitem{OMduality}
C. Montonen and D. Olive, \plb{72}{1977}{117}.\\
P. Goddard, J. Nuyts and D. Olive, \npb{125}{1977}{1}.\\
E. Witten and D. Olive, \plb{78}{1978}{97}.\\
H. Osborn, \plb{83}{1979}{321}.

\bibitem{sw9408} N. Seiberg and E. Witten, 
\npb{431}{1994}{484} [\hepth{9408099}].

\bibitem{aps9505} P.C. Argyres, M.R. Plesser and A.D. Shapere,
\prl{75}{1995}{1699} [\hepth{9505100}].

\bibitem{ab98} P.C. Argyres and A. Buchel, unpublished.

\bibitem{mn9608} J. Minahan and D. Nemeschansky,
\npb{482}{1996}{142} [\hepth{9608047}];
\npb{489}{1997}{24} [\hepth{9610076}].

\bibitem{excft} 
N. Seiberg, \plb{388}{1996}{753} [\hepth{9608111}].\\
O. Ganor, \npb{488}{1997}{223} [\hepth{9608109}].\\
O. Ganor and A. Hanany, \npb{474}{1996}{122} [\hepth{9602120}].\\
N. Seiberg and E. Witten, \npb{471}{1996}{121} [\hepth{9603003}].\\
N. Seiberg, \plb{390}{1997}{169} [\hepth{9609161}].

\bibitem{asty9611} O. Aharony, J. Sonnenschein, S. Yankielowicz and S. Theisen,
\npb{493}{1997}{177} [\hepth{9611222}]. 

\bibitem{dls9612} M.R. Douglas, D. Lowe and J. Schwarz,
\plb{394}{1997}{297} [\hepth{9612062}]. 

\bibitem{a9706} P.C. Argyres, 
\atmp{2}{1998}{293} [\hepth{9706095}].

\bibitem{ab9910} P.C. Argyres and A. Buchel,
\jhep{9911}{1999}{014} [\hepth{9910125}]. 

\bibitem{w9703} E. Witten, 
\npb{500}{1997}{3} [\hepth{9703166}]. 

\bibitem{tHooft} G.~'t Hooft,
``Naturalness, chiral symmetry, and spontaneous chiral symmetry breaking,''
in {\sl Recent Developments in Gauge Theories,} Carg\`ese 1979,
eds.\ G. 't Hooft {\it et.\ al.}\ (Plenum, 1990).

\bibitem{bgisw0507} 
E. Barnes, E. Gorbatov, K. Intriligator, M. Sudano and J. Wright,
\npb{730}{2005}{210} [\hepth{0507137}].

\bibitem{minwalla} S. Minwalla, \atmp{2}{1998}{781} [\hepth{9712074}].

\bibitem{as9509} P.C. Argyres and A.D. Shapere,
\npb{461}{1996}{437} [\hepth{9509175}].  


\bibitem{lll9805} K. Landsteiner, E. Lopez, and D. Lowe,
\jhep{9807}{1998}{011} [\hepth{9805158}].  

\bibitem{acsw0504} P.C. Argyres, M. Crescimanno, A.D. Shapere and 
J. Wittig, ``Classification of N=2 superconformal field theories
with two-dimensional Coulomb branches,"
[\hepth{0504070}]. 

\bibitem{nty9903}
M. Noguchi, S. Terashima and S. Yang, \npb{556}{1999}{115} 
[\hepth{9903215}].

\bibitem{lw90}
W. Lerche and N. Warner, \npb{358}{1991}{571}.

\bibitem{patera}
W. McKay and J. Patera,
{\sl Tables of Dimensions, Indices, and Branching Rules for
Representations of Simple Lie Algebras} (Marcel Dekker, 1981).

\end{thebibliography}
\end{document}